\documentclass[a4paper,11pt]{article}
\pdfoutput=1 

\usepackage{jcappub} 

\usepackage[T1]{fontenc} 
\usepackage{mathtools}

\subheader{\footnotesize
        \vspace*{-3.7em}
        \begin{flushright}
                CERN-TH-2018-176 \\
                MITP/18-071 \\
                INR-TH-2018-018
        \end{flushright}
}

\usepackage{xcolor}
\usepackage{cleveref}

\definecolor{darkgreen}{rgb}{0,0.5,0}

\newcommand{\beq}{\begin{eqnarray}}
\newcommand{\eeq}{\end{eqnarray}}

\title{Femtolensing by Dark Matter Revisited}

\author[a,b]{Andrey~Katz,}
\author[a,c]{Joachim~Kopp,}
\author[a,d,e]{Sergey~Sibiryakov}
\author[a]{and Wei~Xue}

\affiliation[a]{Theoretical Physics Department, CERN, Geneva, Switzerland}
\affiliation[b]{D\'epartement de Physique Th\'eorique and
                Center for Astroparticle Physics (CAP),\\
                Universit\'e de Gen\`eve, 24 quai Ansermet, CH-1211 Gen\`eve 4, Switzerland}
\affiliation[c]{PRISMA Cluster of Excellence \& Mainz Institute for Theoretical Physics, \\
                Johannes Gutenberg University, Staudingerweg 7, 55099 Mainz, Germany}
\affiliation[d]{Institute of Physics, Laboratory for Particle Physics and Cosmology (LPPC),\\ 
            Ecole Polytechnique F\'ed\'erale de Lausanne (EPFL), 
            CH-1015 Lausanne, Switzerland}
\affiliation[e]{Institute for Nuclear Research of the Russian Academy of Sciences, \\
                60th October Anniversary Prospect, 7a, 117312 Moscow, Russia}

\abstract{Femtolensing of gamma ray bursts (GRBs) has been put forward as an exciting
  possibility to probe exotic astrophysical objects with masses below
  $10^{-13}$ solar masses such as small primordial black holes or ultra-compact
  dark matter minihalos, made up for instance of QCD axions.  In this paper we
  critically review this idea, properly taking into account the extended nature
  of the source as well as wave optics effects. We demonstrate that most GRBs
  are inappropriate for femtolensing searches due to their large sizes. This
  removes the previous femtolensing bounds on primordial black holes, implying
  that vast regions of parameter space for primordial black hole dark matter
  are not robustly constrained.  Still, we entertain the possibility that a
  small fraction of GRBs, characterized by fast variability can have smaller
  sizes and be useful.  However, a large number of such bursts would need to be
  observed to achieve meaningful constraints. We study the sensitivity of
  future observations as a function of the number of detected GRBs and of the
  size of the emission region.}

\begin{document}
\maketitle
\flushbottom

\section{Introduction}
\label{sec:intro}

The possibility that the dark matter (DM) in the Universe is made up of massive
compact objects has been explored since the early days of the DM hunt. These
searches have culminated in strong exclusion limits from the
MACHO~\cite{Allsman:2000kg}, EROS~\cite{Tisserand:2006zx},
OGLE~\cite{Wyrzykowski:2011tr} and HSC/Subaru~\cite{Niikura:2017zjd} surveys.
Recently, interest in compact DM objects has been revived, with the most
popular candidates being primordial black holes
(PBHs)~\cite{Carr:1974nx,Clesse:2015wea,Carr:2016drx} or ultra-compact
mini-halos (UCMHs) that are present for example in axion
models~\cite{Hogan:1988mp,Kolb:1993zz,Kolb:1993hw,Hardy:2016mns}.

While a wide range of experimental observations constrain compact objects with
masses above $\sim 10^{-11} \, M_\odot$ (see~\cite{Sasaki:2018dmp} for a recent
review and summary of constraints), bounds on lighter objects are scarcer.
Specifically for PBHs, the requirement that they should not evaporate within
the lifetime of the Universe by emitting Hawking radiation sets a lower bound
on their mass of about $10^{-18}M_\odot$. Moreover, the effect of their
evaporation on Big Bang Nucleosynthesis and the extragalactic photon background
become important already at PBH masses of order $10^{-17}\ M_\odot$, severely
constraining this possibility~\cite{Carr:2009jm}. Other bounds are based on the
existence of old neutron stars \cite{Capela:2012jz, Capela:2013yf,
Pani:2014rca, Capela:2014qea} and white dwarfs~\cite{Graham:2015apa} that would
be destroyed if PBHs were abundant.  Note that there is some controversy in the
literature about the viability of neutron star constraints.  While
ref.~\cite{Pani:2014rca} claims to identify a very efficient mechanism of PBH
capture by neutron stars, the derivation was questioned in
Refs.~\cite{Capela:2014qea,Defillon:2014wla}.  On the other hand the
constraints of ref.~\cite{Capela:2013yf} hinge on the assumption that the
globular clusters hosting neutron stars are embedded in overdense DM cores.
While this  assumption is not ruled out, it is also not experimentally
supported~\cite{Conroy:2010bs,Ibata:2012eq}. Indeed, it has even been shown
that globular clusters could be formed without any DM
overdensities~\cite{Naoz:2014bqa,Popa:2015lkr}. 

None of the bounds discussed above applies directly to UCMHs, which
below $10^{-11}M_\odot$ remain essentially unconstrained.

To probe the mass range between $\sim 10^{-17} \, M_\odot$ and $\sim 10^{-13}
\, M_\odot$, femtolensing of distant gamma-ray bursts (GRBs) has been proposed
by Gould~\cite{Gould:1991td} (see also ref.~\cite{Nemiroff:1995ak} for a
proposal along the same lines for somewhat higher masses).  The basic idea
behind femtolensing is that, while the two images of the GRB created by such a
tiny lens cannot be resolved in space or time, their wave fronts will acquire
different phases during propagation because of the different path lengths and
gravitational potentials they experience.  If the phase shift is of order one,
one expects to see interference fringes in the frequency spectrum.  Since
setting meaningful limits with this method requires a sufficient number of
GRBs, most works on femtolensing have been purely theoretical until very
recently.  An early experimental analysis based on BATSE data was
presented in ref.~\cite{Marani:1998sh}, but yielded only very weak bounds.  The
situation has changed dramatically with the advent of Fermi data. There are now
dozens of GRBs with reliably measured redshift and frequency spectrum, with
hopes to significantly expand this sample in the near future.  Fermi data has
prompted first observational works on femtolensing of GRBs by black holes: The
authors of ref.~\cite{Barnacka:2012bm} claim to constrain primordial black
holes with masses between $10^{-16}$ and $10^{-14} \, M_\odot$ to contribute no
more than 10\% to the total dark matter abundance in the Universe.  Although
the validity of these bounds has been questioned later because they are based
on the assumption that the gamma-ray source is point-like in the lens
plain~\cite{Pani:2014rca,Davidson:2016uok} (see also the relevant estimations
of~\cite{Barnacka:2014yja}), they have been widely accepted as the state of the
art.

In this work we critically re-analyze the idea of GRB femtolensing.  We will
use the techniques of~\cite{1993ApJ...413L...7S}, further elaborated
in~\cite{Matsunaga:2006uc}.  One of the questions we study is the impact of the
non-pointlike nature of GRBs.  We will see that most GRBs are too big when
projected onto the lens plane to yield meaningful femtolensing limits. Only a
small population of GRBs with very fast variability might be suitable for such
searches, but even their size is comparable to the Einstein radius of the
lenses of interest to us. It must therefore be properly taken into account.

A second question we address is to what extent femtolensing can be used to
constrain non-point-like lenses, i.e., objects that extend beyond their
Einstein radius. An important example are UCMHs consisting of DM.  Indeed, it
has been noticed in the 1990s that femtolensing might be relevant for probing
UCMHs in the mass range between $10^{-16}$ and $10^{-12}$ solar
masses~\cite{Kolb:1995bu}.  But this reference still treated the UCMHs as
point-like, which is not necessarily in agreement with more recent estimates of
their size~\cite{Zurek:2006sy}.  Femtolensing of UCMHs has not been re-analyzed
since then, in spite of the significant progress that has been made both in our
theoretical understanding of compact DM objects and in GRB observations.
 
Besides accounting for the finite size of the
source and lens, we will also emphasize that the geometric optics approximation
used in the original proposal~\cite{Gould:1991td} breaks down exactly in the
mass range where femtolensing appears most promising.  In other words,
femtolensing in this mass range cannot be described by considering just two
images of the source, with properties derived from the geometry of the
corresponding lines of sight.  Originally, the importance of wave optics
effects in femtolensing had been pointed out in \cite{Ulmer:1994ij}.

We will find that, especially because of the non-negligible size of the source, the
data available to date does not constrain primordial black holes or other
compact DM objects. We also compute the sensitivity of hypothetical future
surveys as a function of the number of observed GRBs and their transverse size.

The paper is organized as follow. In \cref{sec:theory} we describe the theory
of femtolensing, taking into the account all the caveats discussed above:
finite size of the source, possibly finite size of the lens, breakdown of the
geometric optics approximation. In \cref{sec:bounds}, we review the
current femtolensing bounds on compact DM objects and show that they do not
cover the physically interesting parameter region. We also show projections
into the future, arguing that a sample of 100 GRBs with transverse size
$10^9\,\text{cm}$ would be needed to exclude DM in the form of PBHs in the mass
range from $10^{-16}$ to $5\times 10^{-15}$ solar masses.  We conclude in
Sec.~\ref{sec:conclusions}. Details on the size of the emission region in a GRB
are relegated to the Appendix.

\section{Theory of Femtolensing}
\label{sec:theory}

In this section we review the general idea of femtolensing and the underlying
formalism, and we address several important caveats.  We will start by
discussing the case of a point-like lens affecting light from a point-like
source in the geometric optics approximation. We then introduce one-by-one the
wave optics corrections, the effect of an extended source, and the possibility
of an extended lens whose size exceeds its would-be Einstein radius.  To the
best of our knowledge, this study has not been performed before, but is
strongly motivated by the appeal of ultra-compact DM miniclusters.

\subsection{Point-like Lens and Source, Geometric Optics Regime}
\label{sec:basic-formalism}

The basic femtolensing scenario, put forward in~\cite{Gould:1991td}, is based
on the assumption that a gamma ray emitted by a point-like source with a
non-zero impact parameter with respect to the lens--observer axis, is split by
the lens into two rays, each of which is delayed with respect to the unlensed
case by some time shift $\Delta t_i$ ($i=1,2$).  This corresponds to a phase
shift of $\Delta \phi_i \equiv \omega \Delta t_i$, where $\omega$ is the
angular frequency of the photons.  If the two images cannot be resolved in space
and time, the two rays will interfere,
producing characteristic fringes in the spectrum. 

In the thin lens approximation the time delay is given
by~\cite{Bartelmann:2010fz}  
\begin{align}
  \Delta t = \frac{1}{c} \frac{D_L D_S}{D_{LS}} (1 + z_L) 
             \left( \frac{|\vec \theta - \vec \beta|^2}{2} - \psi (\vec \theta)  \right) \,.
  \label{eq:timedelay}
\end{align}
Here $D_L$, $D_S$, and $D_{LS}$ are
the angular diameter distances between the observer and the lens, the observer
and the source, and the lens and the source, respectively.  The redshift of the
lens is denoted $z_L$, while $\beta$ is the angle under which the observer
would see the source in the absence of a lens, and $\theta$ is the angle under
which the observer sees a given point in the lens plane. The function
$\psi(\theta)$ is the lensing potential which is related to the density profile
$\rho(r)$ of the lens by the
Poisson equation
\begin{align}
  \nabla^2 \psi(\theta)
    \equiv \frac{1}{\theta} \frac{\partial}{\partial\theta}
           \bigg( \theta \frac{\partial \psi}{\partial\theta} \bigg)
    = \frac{8\pi G}{c^2} \frac{D_{LS} D_L}{D_S}
      \int_{-\infty}^\infty \! d\xi \, \rho\big(\sqrt{(D_L \theta)^2 + \xi^2} \big) \,.
  \label{eq:lensing-potential}
\end{align}
Here, the integral runs along the line of sight, and $\nabla^2$ is the
two-dimensional Laplace operator, which we express in polar coordinates, with
$\theta$ being the radial direction. We have also assumed a spherically symmetric
lens.  For a point-like lens of mass $M$,\footnote{Note that the lensing
  potential is defined by \cref{eq:lensing-potential} only up to an
  additive constant. We
  will ignore this constant here, which implies that the expression for the time delay,
  \cref{eq:timedelay}, is applicable only for calculating the time difference
  between different paths, but does nor necessary reflect the \emph{absolute}
time delay along the path.}
\begin{align} 
  \psi(\theta) &= \theta_E^2 \log\theta
               &  \text{(point-like lens)} \,,
\end{align}
where
\begin{align}
  \theta_E \equiv \left( \frac{4 G M}{c^2} \frac{D_{LS}}{D_S D_L} \right)^{1/2} \, 
  \label{eq:theta-E}
\end{align}
is the Einstein angle (i.e.\ the Einstein radius $R_E$ divided by $D_L$).  On
top of being a convenient definition, the Einstein angle has a well
defined physical meaning: in the geometric optics approximation,
it is the size of the Einstein ring produced by
a point-like source aligned with the observer--lens line of sight ($\beta=0$).
For the more general density profiles that we will discuss in
\cref{sec:extended-lens} the location of the ring will be different from
$\theta_E$ and will be called $\theta_0$.  For point-like masses, the two
values coincide.  

In geometric optics, Fermat's principle stipulates that the images will be seen
under those angles $\theta$ for which $\Delta t$ is stationary.  For point-like
lens and source, this requirement leads to the lens equation\footnote{The generic 
lens equation is $\vec \theta - \vec \beta = \vec \nabla \psi(\vec \theta) $.}
\begin{align}
  \theta - \beta = \frac{\theta_E^2}{\theta} \,.
  \label{eq:lens-eq}
\end{align}
Lensing is typically observable when $\theta \sim \theta_E$ (or, more generally, 
when $\theta\sim \theta_0$ for non-point-like lenses), otherwise one
of the images becomes extremely faint. This shows that both terms in
\cref{eq:timedelay} scale as $\theta_E$ up to order-one factors. The time delay
thus depends only on the mass of the lens and is practically insensitive to the
distances either to the source or to the lens: $\Delta t \sim 4 G M/c^3 =
2 R_s / c$, where $R_s$ is the Schwarzschild radius of the lens.
Assuming that the source emits gamma rays in the $10$ to $1000$~keV range,
we see that order-one phase shifts occur for lens masses between
$10^{-17}$ and $10^{-14}$ solar masses. 
For lensing at cosmological distances $\sim 1$\,Gpc the corresponding
Einstein angles fall in the femto-arc-second range, which explains the
term ``femtolensing''.

The solutions to the lens equation (\ref{eq:lens-eq}) are
\begin{align}
  \theta_\pm = \frac{1}{2} \left( \beta \pm \sqrt{\beta^2 + 4 \theta_E^2} \right) \,.
\end{align}
The magnifications of the two images are given by~\cite{Bartelmann:2010fz}
\begin{align}
  \mu_\pm = \frac{y^2 + 2}{2 y \sqrt{y^2 + 4}} \pm \frac{1}{2} \,,
  \label{eq:magnification-geom-1}
\end{align}
where we have defined 
\begin{align}
  y \equiv \beta / \theta_E\;.
  \label{ydef}
\end{align}
Taking into account the phase shift between the images, the total intensity is
proportional to~\cite{Matsunaga:2006uc}
\begin{align}
  \mu = \frac{y^2 +2}{y \sqrt{y^2 + 4}} + \frac{2}{y \sqrt{y^2 + 4}}
        \sin \left(\Omega
               \left[ \frac{y \sqrt{y^2 + 4}}{2}
                          + \log \left| \frac{y + \sqrt{y^2 + 4}}{y - \sqrt{y^2 + 4}} \right|
               \right] \right) \,,
  \label{eq:magnification-geom-2}
\end{align}
where we have introduced the dimensionless frequency as
\begin{align}
  \Omega \equiv \frac{1}{c}\frac{D_S D_L }{D_{LS}} \theta_0^2 (1 + z_L ) \, \omega
  \equiv \Delta t_0 \, \omega \,.
  \label{eq:Omega_def_gen}
\end{align}
Here we have also introduced a new quantity $\Delta t_0 $ which is a typical 
time delay between different images in the geometric optics picture. 
Note that this definition of the dimensionless frequency is completely generic and we
are going to use it also for extended mass distributions 
in \cref{sec:extended-lens}. For point-like lenses, however, 
\cref{eq:Omega_def_gen} simplifies to 
\begin{align}
  \Omega \equiv \frac{4 G M (1 + z_L)}{c^3}\,\omega \,.
  \label{eq:Omega_def}
\end{align}
\Cref{eq:magnification-geom-2} clearly depends on the photon energy $\omega$ and
varies between a maximum and minimum value as a function of $\omega$.
Therefore a signal of femtolensing would be an oscillatory pattern of
interference 
fringes in the otherwise smooth GRB spectrum. 

Let us list the conditions required for the validity of the above formulas.
\begin{enumerate}
  \item The limit of geometric optics should be
    applicable.\footnote{We are grateful to  Juan Garcia-Bellido for pointing
    out to us the importance of wave optics effects.} This is the case as long
    as
    \begin{align}
      \omega \, \Delta t_0 \gg 1
      \label{eq:geomopt}
    \end{align}
    (see \cref{sec:wave-optics} for details).  In the case of a point-like lens,
    this reduces (up to a factor $4 \pi (1+z_L)$) to the condition that the
    photon wave length should be much smaller than the Schwarzschild radius
    $R_s$ of the lens.  Note, however, that this \emph{does not mean} that the
    Schwarzschild radius can be interpreted as the effective radius of the
    lens, because the typical deflection distance of the photons is $R_E$
    rather than $R_s$. Moreover, the relation \cref{eq:geomopt} to the
    Schwarzschild radius is something very peculiar to point-like lenses and
    does not hold for more general mass distributions.  A simple estimate shows
    that the inequality \eqref{eq:geomopt} breaks down for $M \lesssim 10^{-15}
    M_\odot$ for a gamma ray energy of $E \sim 100$~keV, essentially
    invalidating the geometric optics approach in most of the parameter space
    that is relevant for femtolensing. The breakdown of the geometric optics
    approximation \emph{does not} imply the absence of an observable
    femtolensing effect. It does, however, mean that a full wave optics
    computation is necessary (see \cref{sec:wave-optics}).
  
  \item The source should be point-like in the plane of the lens.  The signals
    from different points of the source add up incoherently, and the condition
    that this does not wash out the interference fringes is
    \begin{align}
      \sigma_y \lesssim \frac{1}{\Omega}\;,
      \label{Deltay}
    \end{align}
    where $\sigma_y$ is the angular size of the source normalized to
    $\theta_0$.
    This condition can be easily derived from \cref{eq:magnification-geom-2}.
    Since interference fringes can only be observed if $\Omega \gtrsim 1$,
    condition \eqref{Deltay} also implies that the size of the
    source projected on the lens plane should be smaller than the Einstein radius
    for point-like lenses ($\theta_0$ for extended lenses).
    As we will see later, this condition is never fully satisfied for
    femtolensing of GRBs. At best, the projected size of the source can be
    comparable to the Einstein radius, so that finite size effects can never be
    neglected. Note also that the constraint on the size of the source is
    stricter for higher frequencies/energies. Therefore, as the size
    of the source increases, the interference fringes will first
    disappear at high frequencies. 

  \item The source emission must be coherent over timescale of the time
    delay, \cref{eq:timedelay}.  Usually this condition is easily
    satisfied because it merely requires a detector with sufficient energy
    resolution.  Typically, we are interested in scenarios where
    $\omega \, \Delta t_0 \sim 1$. Given that any reasonable detector has
    an energy (or frequency) uncertainty $\delta\omega \ll \omega$, the
    Heisenberg uncertainty principle tells us that the coherence time 
    $\delta t \sim \delta\omega^{-1}$ is much larger than $\Delta t_0$.
\end{enumerate} 
Because the conditions~(1) and~(2) are never strictly satisfied, we
are forced to relax some of the approximations made above. We will do so
in the following subsections.

\subsection{Point-like Source and Lens in the Wave Optics Regime}
\label{sec:wave-optics}

Gravitational lensing outside the geometric optics regime is not an unusual
scenario. For example, it has been mentioned in~\cite{Inomata:2017vxo} that the
microlensing measurements of HSC Subaru are partially in the physical (wave)
optics regime, invalidating the constraints on primordial black holes published
in ref.~\cite{Niikura:2017zjd} for masses below $10^{-11} M_\odot$.  To the
best of our knowledge, the correct interpretation of the HSC constraints in
this mass range is still missing in the literature.  Also, practically all
discussion of lensing of gravitational waves includes wave optics
effects~\cite{Takahashi:2003ix} (for recent related works see
e.g.~\cite{Jung:2017flg,Christian:2018vsi,Li:2018prc}).  Finally, it was
noticed already in the 1990s that the geometric optics approximation is
typically violated in femtolensing~\cite{Ulmer:1994ij}.   

\begin{figure}[t]
  \centering 
  \includegraphics[width=.48\textwidth]{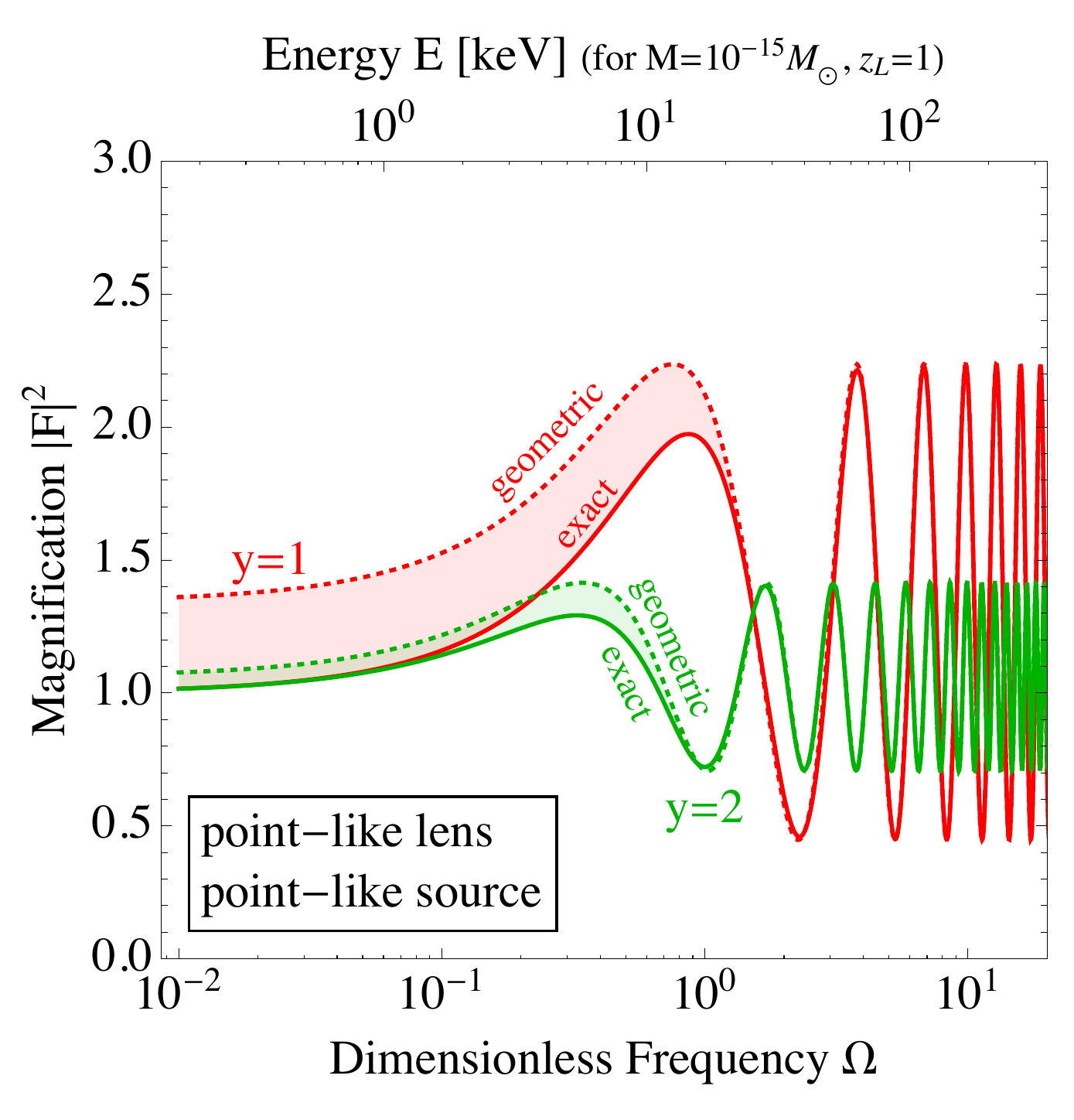} 
  \caption{Comparison of the interference picture in the geometric
    optics approximation (dashed) and the full wave optics results (solid) in
    the case of a point-like source and a point-like lens.  Wave effects
    strongly impact the  magnification of the signal at low energies, while at
    higher energies, geometric optics provides a good approximation.  The
    magnification is plotted against the dimensionless frequency  $\Omega$
    defined in Eq.~\eqref{eq:Omega_def}.}
  \label{fig:BH_phys_vs_geom}
\end{figure}

In the physical optics regime, lensing is characterized by a magnification
function $F(\vec y;\omega)$, which is defined as the ratio $\phi_L / \phi$,
where $\phi_L$ and $\phi$ are the electromagnetic wave amplitudes with and
without lensing, respectively.  The magnification of the signal intensity is
therefore 
\begin{align}
  \mu = |F|^2 \,,
  \label{eq:magnification-wave}
\end{align}
The
magnification function is given by~\cite{Nakamura:review}
\begin{align}
  F(\vec y;\omega) = \frac{\Omega}{2 \pi i}
         \int \! d^2\vec{x}\; e^{i \omega \Delta t(\vec x, \vec y)} \,.
  \label{eq:physopt}
\end{align}
where the dimensionless frequency $\Omega$ has been defined in
\cref{eq:Omega_def_gen}, $x \equiv \theta / \theta_0$, and the time delay $\Delta t(\vec{x},
\vec{y})$ is taken from \cref{eq:timedelay}.   The integral thus runs over the
lens plane, with each point in the plane contributing to $F(\vec{y},\omega)$.
For a point-like lens like a black hole, the integral can be evaluated
analytically, leading to
\begin{align}
  F(y, \Omega)_\text{BH} = e^{i \Omega |\vec y|^2/2} \,  
    \left(-\frac{ i \Omega}{2}\right)^{i \Omega/2} \,
    \Gamma \left( 1 -\frac{i \Omega}{2} \right)
    L_{-1 + \frac{i \Omega}{2}} \left(-\frac{i |\vec y|^2 \Omega}{2} \right) \,,
  \label{eq:Fy}
\end{align}
where $L_n$ is the $n$-th order Laguerre polynomial. For an
extended lens, analytic expressions in general do not exist, so the integral in
\cref{eq:physopt} must be evaluated numerically.

The geometric optics regime corresponds to the saddle point approximation
of~\eqref{eq:physopt}: as discussed in \cref{sec:basic-formalism}, the lens
equation (\ref{eq:lens-eq}) is obtained by imposing the stationary point
condition $\vec \nabla_x \Delta t = 0$.  Also the magnification $\mu$ in the
geometric optics limits, \cref{eq:magnification-geom-1}, can
be derived from the saddle point method. However, the saddle point method
only provides a good approximation to the full integral if the phase is large,
which explains why geometric optics only works if $\omega\, \Delta t \gg 1$.

In Fig.~\ref{fig:BH_phys_vs_geom}, we compare the intensity calculated using the
exact wave optics expression, \cref{eq:physopt}, to the one obtained in the
geometric optics approximation, \cref{eq:magnification-geom-2}.
We see that the geometric optics approximation overestimates the amplitude
of the first fringe in the spectrum, while at higher photon energies, it
is applicable.  We will see in \cref{sec:extended-lens}
that the effects of wave optics are more important for extended lens profiles.

\subsection{Extended Sources}
\label{sec:extended-source}

As argued above, interference patterns due to femtolensing are only observable
if the size of the source projected onto the lens plane is not
much larger than 
the Einstein radius of the lens (assumed to be point-like in this subsection).
In other words, we require that
\begin{align}
  \sigma_y \equiv \frac{a_S}{D_S \theta_E} \ll 1 \,,
  \label{sigmay}
\end{align}
where $a_S$ is the actual (unprojected) transverse size of the emission region.
Most GRBs have an $\mathcal{O}(1)$ redshift, corresponding to
$D_S \sim \text{Gpc}$. Therefore a lens of mass $M \sim 10^{-15} M_\odot$
that is also located at $z_L \sim \mathcal{O}(1)$  has an Einstein radius
of order $R_E \sim 10^{9}$~cm. 
The Einstein radius of lighter lenses will be even smaller.
Unfortunately, the emission size of most GRBs is much
larger than this \cite{Barnacka:2014yja,Golkhou:2015lsa}.

As discussed in \cref{app:GRBsize}, the majority of GRBs are
likely to have transverse sizes of about $a_S \sim 10^{11}\,\text{cm}$,
which is two orders of magnitude larger than required for observable
femtolensing. Still, the possibility that a small fraction of GRBs
have much smaller sizes $a_S\lesssim 10^9 \, \text{cm}$ is not
excluded. We will entertain this possibility and study the dependence
of the femtolensing signal on the size of the source. Hopefully,
future developments in GRB modeling, perhaps on a case-by-case
basis, will lead to more precise estimates of their sizes than
are currently available, and will allow us to cherry-pick events suitable
for femtolensing. Note that, if such GRBs exist, they will be
characterized by very fast intrinsic variability at time
  scales\footnote{We leave aside the question whether such variability
  time scales can be measured with existing instruments, as the main
  focus of our study are future observations.} $t_{\rm var}\lesssim 
0.5\times10^{-3}$ sec.  

To take the finite size of the source into account,
we follow the formalism of ref.~\cite{1993ApJ...413L...7S}.  The observed
magnification is then
\begin{align}
  \bar\mu = \frac{\int \! d^2y \, W(\vec y; \sigma_y) \, \mu(\vec y; \Omega)}
                 {\int \! d^2y \, W(\vec y; \sigma_y)} \,,
\end{align}
where $W(y; \sigma_y)$ is a window function that describes the
intensity profile 
of the emission 
and $\mu(\vec y; \Omega)$ is the magnification for a point-like
source, see eqs.~(\ref{eq:magnification-wave}), (\ref{eq:physopt}).
In principle $W(\vec y)$ can be any well-behaved function that acts as a mask of
size $\sigma_y$. We choose it to be a Gaussian,
\begin{align}
  W(y; \sigma_y) = e^{-|\vec y - \vec y_0|^2/2\sigma_y^2} \,.
\end{align}
Hereafter, we will use $\vec y_0$ to denote the location of the center of the
emission.  For Gaussian $W(y; \sigma_y)$ and a radially symmetric lens, the
weighted magnification reads
\begin{align}
\label{mu-extended-source}
  \bar\mu = \frac{e^{-y_0^2/2\sigma_y^2}}{\sigma_y^2} \int_0^\infty \!
    dy \, y \, e^{-y^2/2\sigma_y^2}\, 
    I_0 \bigg( \frac{y_0\, y}{\sigma_y^2} \bigg) \, \mu(y; \Omega)
\end{align}
where $I_0(x)$ is the modified Bessel function of the first kind.

\begin{figure}[t]
  \centering 
  \includegraphics[width=.48\textwidth]{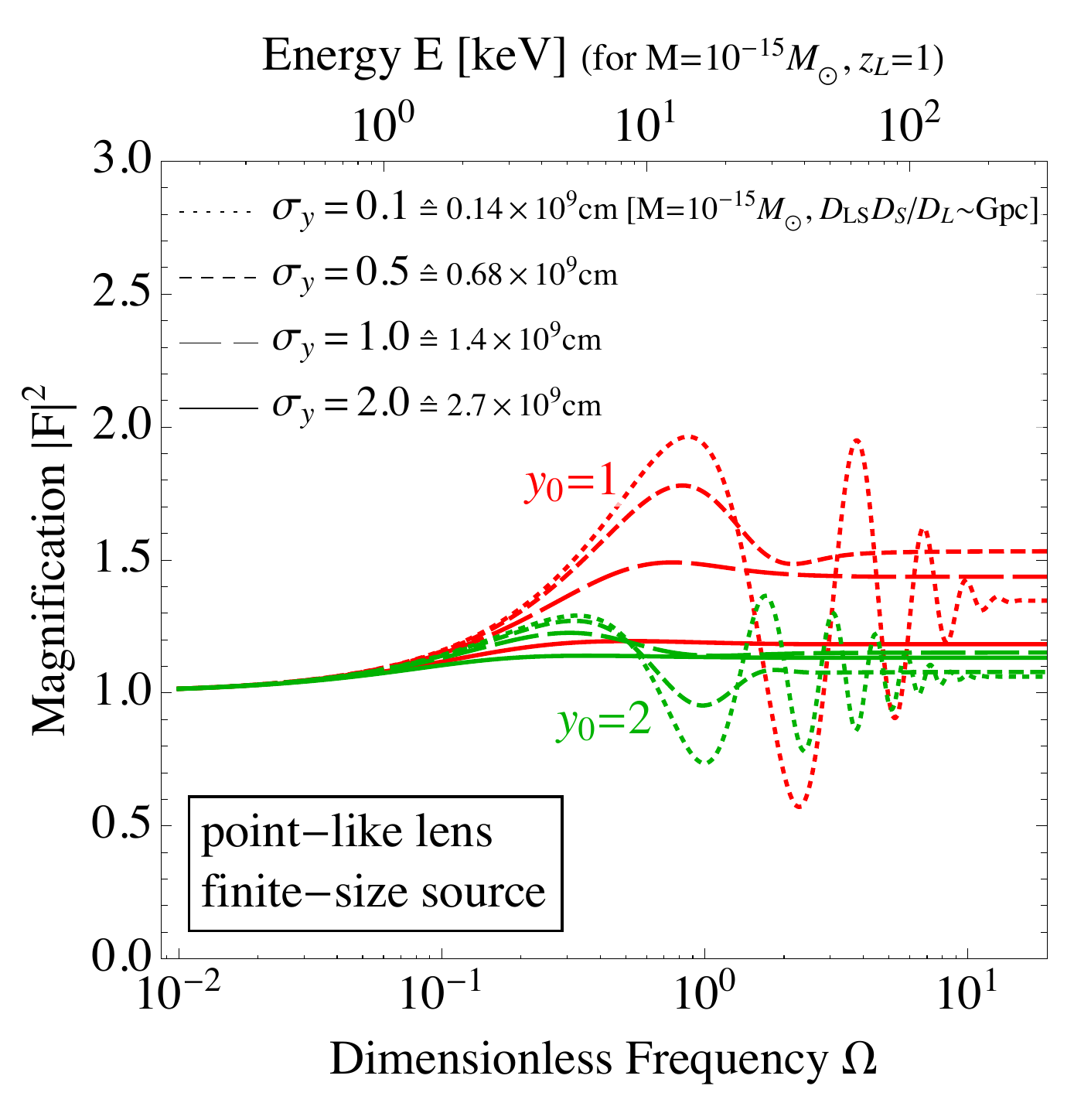}
  \caption{ Dependence of magnification on the size of the source. Here
    $\sigma_y$ is the angular size of the source in units of the Einstein
    angle, see \cref{sigmay}. The dimensionless frequency $\Omega$ is
    defined in \cref{eq:Omega_def}. The lens is assumed to be point-like.}
  \label{fig:BH_phys_extended}
\end{figure}

We illustrate the effect of the non-zero source size in
\cref{fig:BH_phys_extended}.  Clearly, the effect is mild for $\sigma_y \ll 1$,
however even for these values of the emission size the oscillations in energy
space are damped at high frequencies. As the emission size grows, oscillations
are damped more strongly, until they eventually disappear at $\sigma_y \sim 1$. 

Interestingly, we observe that, even at large emission size $\sigma_y\gtrsim 1$,
the asymptotic $\bar\mu$ at high frequency is larger than the value of $\bar\mu$
at low frequency. That is, even though the interference fringes are washed out,
a smooth step-like feature survives.  This can be understood as follows. At
$\Omega \to \infty$ the point-source magnification $\mu(y;\Omega)$ is
well described by the geometric optics expression
(\ref{eq:magnification-geom-2}). It quickly oscillates as a function
of $y$ and can be replaced in the integral (\ref{mu-extended-source})
by its mean value,
\begin{align}  
  \mu(y; \Omega) \mapsto \frac{y^2+2}{y\sqrt{y^2+4}} \,.
\end{align}
Thus, in the high frequency limit the weighted magnification reads  
\begin{align}
  \lim_{\Omega\to\infty}\bar\mu=
  \frac{e^{-y_0^2/2\sigma_y^2}}{\sigma_y^2} \int_0^\infty \!
  dy \, \frac{y^2+2}{\sqrt{y^2+4}} \, e^{-y^2/2\sigma_y^2}\, 
            I_0 \bigg( \frac{y_0\, y}{\sigma_y^2} \bigg) > 1\;.
\end{align}
On the other hand, in the limit $\Omega \to 0$ the geometric optics
approximation is not applicable and the wave optics calculation leads to
$\lim_{\Omega \to 0} \bar\mu = 1$. The difference between the two limiting
values leads to the step-like feature in the weighted magnification function at
low dimensionless frequencies observed in \cref{fig:BH_phys_extended}.  Of
course, in the absence of oscillatory fringes, one probably cannot rely on this
feature to establish that a given GRB spectrum has been lensed. On the other
hand, a non-observation of such feature can potentially be used to
\emph{exclude} lensing.  Since our proposal to perform an exclusion based on
the absence of this feature is rather speculative, we will not use it in
estimating the future reach in \cref{sec:bounds}. However, if proven possible,
this would imply that even $\gamma$-ray sources as big as $10^{10}$\;cm might be
used to exclude femtolensing.

\subsection{Extended Lens}
\label{sec:extended-lens}

We finally analyze the effects of a non-point-like lens, namely a lens whose
extent is larger than its would-be Einstein radius.  A notable example of such
lenses would be ultracompact minihalos (UCMHs) composed of DM. These are
predicted, for instance, in scenarios that involve DM in the form of QCD axions
if the Peccei--Quinn phase transition happens after inflation. In this case,
overdensities that arise after the phase transition due to different initial
values of the axion field in different Hubble patches collapse into UCMHs
around matter--radiation equality~\cite{Hogan:1988mp,
Kolb:1993hw,Zurek:2006sy}.  The mass and radius estimates of UCMHs vary widely
in the literature. For example, an average  value for the mass of the QCD axion
miniclusters as small as $10^{-14}~M_\odot$ was suggested
in~\cite{Enander:2017ogx}.  This is two orders of magnitude lighter than
the previous estimates
of~\cite{Kolb:1995bu,Tinyakov:2015cgg,Davidson:2016uok,Bai:2016wpg}.  Even
bigger masses were suggested in~\cite{Fairbairn:2017dmf,Fairbairn:2017sil}.
Although resolving these discrepancies is beyond the scope of this work, some
part of the suggested parameter space is likely to be within the reach of
femtolensing searches, as has been first pointed out in~\cite{Kolb:1995bu}.
UCMHs are not unique to QCD axions and can be formed also in many other models
with axion-like particles~\cite{Hardy:2016mns}.  The parameter space explored
in ref.~\cite{Hardy:2016mns} is vast, and some of it is definitely accessible
to femtolensing, in particular axion-like particles with
temperature-independent ($n=0$) masses of $\sim 10^{-3}$--$10^{-5}$~eV.

Another parameter subject to significant uncertainties is the
radius of the UCMHs.  However, what is 
important for our discussion is that it is most likely bigger than its 
would-be Einstein radius. For instance, 
for the QCD axion, assuming miniclusters are spherically symmetric, 
refs.~\cite{Kolb:1995bu,Zurek:2006sy,Tinyakov:2015cgg} give,
\begin{align}
  R_\text{UCMH} \simeq \frac{3\times 10^{12} \; \text{cm}}{\Phi(1+\Phi)^{1/3}}
    \left( \frac{M}{10^{-12} M_\odot} \right)^{1/3} \,,
  \label{eq:R-UCMH}
\end{align}
where $\Phi$ is the initial axion density contrast.  Most miniclusters will
have $\Phi\lesssim 10$ \cite{Kolb:1995bu}, so their radius according to
\cref{eq:R-UCMH} is almost two orders of magnitude larger than their
corresponding Einstein radius.  Only for extremely dense miniclusters with
$\Phi\gtrsim 100$, the size and the Einstein radius become comparable.  It is
evident from these estimates that one cannot neglect the extent of the UCMHs in
the lensing problem.

Finally, there is no agreement in the literature on the density profile of
UCMHs, which may also depend on the mechanism by which they form in the early
Universe.  Some scenarios suggest rather steep profiles.  For example, the
self-similar infall scenario motivates a density profile of the form
$\rho(r)\propto r^{-9/4}$ \cite{Fillmore:1984wk, Bertschinger:1985pd}.  This
profile has been confirmed in N-body simulations~\cite{Zurek:2006sy,
Vogelsberger:2009bn, Vogelsberger:2010eh}, but its relevance to UCMHs was later
questioned in~\cite{Delos:2017thv}.  The latter paper advocates more shallow
profiles, $\rho(r) \propto r^{-3/2}$, or the Navarro--Frenk--White (NFW)
profile which scales as $\rho(r) \propto r^{-1}$ at small radii.
Given this uncertainty, we prefer to be agnostic about the precise shape of the
UCMH profile. Rather we will assume that its inner part is described by a
generic power law cusp, 
\begin{align}
  \rho(r) = \rho_0 \left( \frac{r}{r_0}\right)^{-\delta}~,~~~~\delta<3 \,,
  \label{eq:self-similar}
\end{align}
where $r_0$ characterizes the size of the inner part of the UCMH and $\rho_0$
is the density at that  distance.\footnote{Note that some steep profiles tend
  to develop a core at the center of the distribution. In particular, the
  self-similar radial infall profile corresponding to $\delta=9/4$ cannot be a
  valid approximation all the way down to $r = 0$ (see e.g.~\cite{Bringmann:2011ut}
  for an approach to estimate the size of the core).  While we will for
  simplicity neglect the presence of the core in the sensitivity studies
presented in this paper, it should be taken into account when analyzing real
data.} It is convenient to introduce the mass enclosed within radius $r_0$,
which we will call the ``cusp mass'',
\begin{align}
  M_\text{cusp} = \frac{4\pi}{3-\delta}\rho_0 r_0^3 \,.
\end{align}
Note that the total mass of UCMHs can, in general, be bigger than
$M_\text{cusp}$ and can, in fact, even be formally divergent.  The Einstein
radius corresponding to $M_\text{cusp}$ is
\begin{align}
  R_{E,\,\text{cusp}} = \bigg( \frac{4 G M_\text{cusp}}{c^2}
                               \frac{D_{LS} D_L}{D_S} \bigg)^{1/2} \,.
 \label{eq:rE-UCMH}
\end{align}
It is this radius that we mean when referring to the Einstein radius of
the UCMH.  We remind the reader that for extended lenses, the Einstein
radius does not have an immediate physical interpretation.
We assume $R_{E,\,\text{cusp}} < r_0$, such that the cusp cannot
be treated as point-like for the purposes of gravitational lensing.

It is easy to see that only profiles with $\delta>1$ can be relevant
for femtolensing. Indeed, the mass enclosed within a radius $r$ grows
as $M(r) \propto r^{3-\delta}$ and the corresponding Einstein radius
scales as $R_E(r) \propto r^{(3-\delta)/2}$. If $\delta \leq 1$,
any part of the lens is bigger than its Einstein radius, $R_E(r) < r$,
so no multiple images and thus no femtolensing can arise. On the
other hand, for $\delta>1$, the central part of the cusp happens to be
within its Einstein radius. Thus, it acts qualitatively similar to a
point-like lens leading to appearance of multiple images.

We now study the case $\delta>1$ in more detail. 
Following \cref{eq:lensing-potential}, the lensing potential corresponding to
the density profile \cref{eq:self-similar} is
\begin{align}
  \psi(\theta) = \frac{\theta_0^2}{3-\delta}
                  \bigg( \frac{\theta}{\theta_0} \bigg)^{3-\delta}\,,
  \label{eq:psi-self-similar}
\end{align} 
where we have defined
\begin{align}
  \theta_0 = \kappa(\delta)\, \frac{R_{E,\,{\rm cusp}}}{D_L}
         \bigg( \frac{R_{E,\,{\rm cusp}}}{r_0}
         \bigg)^{\frac{3-\delta}{\delta-1}}\,,
  \qquad
  \kappa(\delta)\equiv \bigg(
    \frac{\sqrt{\pi}\,\Gamma\big((\delta-1)/2\big)}{2\,\Gamma(\delta/2)}
  \bigg)^{\frac{1}{\delta-1}} \;.
  \label{eq:theta-0-UCMH}
\end{align}  
The angle $\theta_0$ is an analog of the Einstein angle for extended
lenses. In the geometric optics approximation it coincides with the
angular size of the Einstein ring produced by a point-like source
aligned with the center of the lens ($\beta=0$ in the notations of
section \ref{sec:basic-formalism}). For misaligned sources ($\beta\neq
0$), $\theta_0$ sets the characteristic distance between their
multiple images. The characteristic time delay $\Delta t_0$ is
defined in~\eqref{eq:Omega_def_gen}. 
Note that $\theta_0$ is parametrically smaller than the naive Einstein
angle $R_{E,\,\text{cusp}}/D_L$.

For our subsequent discussion, we need to distinguish between
the cases $\delta < 2$ and $\delta > 2$. For $1 < \delta < 2$ and small
enough $\beta$, the lens equation $\vec
\theta-\nabla\psi(\vec\theta)=\vec\beta$ has three solutions
corresponding to three images in the geometric optics
approximation. Above a certain critical value $\beta_\text{cr}
 = C(\delta) \theta_0$, where $C(\delta)$ is an order-one
coefficient, two of the images disappear and only a single one
remains. In other words, the ring $\beta = \beta_\text{cr}$ is a
caustic. Thus, the appearance of interference fringes characteristic for
femtolensing is possible only if
the source is close enough to the line of sight passing through the
center of the lens. Then, the existence of more than two images will in
general lead to a complicated interference pattern
(cf.~ref.~\cite{Ulmer:1994ij}). 

The properties of a lens with $2<\delta<3$, including the self-similar infall
profile with $\delta = 9/4$, are closer to those of point-like objects like
PBHs. In this case the lens equation always has two solutions corresponding to
two images in the geometric optics approximation. This will give rise to the
characteristic sinusoidal dependence of the magnification on frequency. The
caustic shrinks to the point $\beta=0$, at which the two images turn into an
Einstein ring of angular radius $\theta_0$.\footnote{The case isothermal sphere
profile with $\delta=2$ lies at the boundary between the two regimes discussed
here: there are two lensed images for $\beta<\theta_0$ and a single image for
$\beta>\theta_0$.}
 
\begin{figure}
  \centering
  \includegraphics[width=0.48\textwidth]{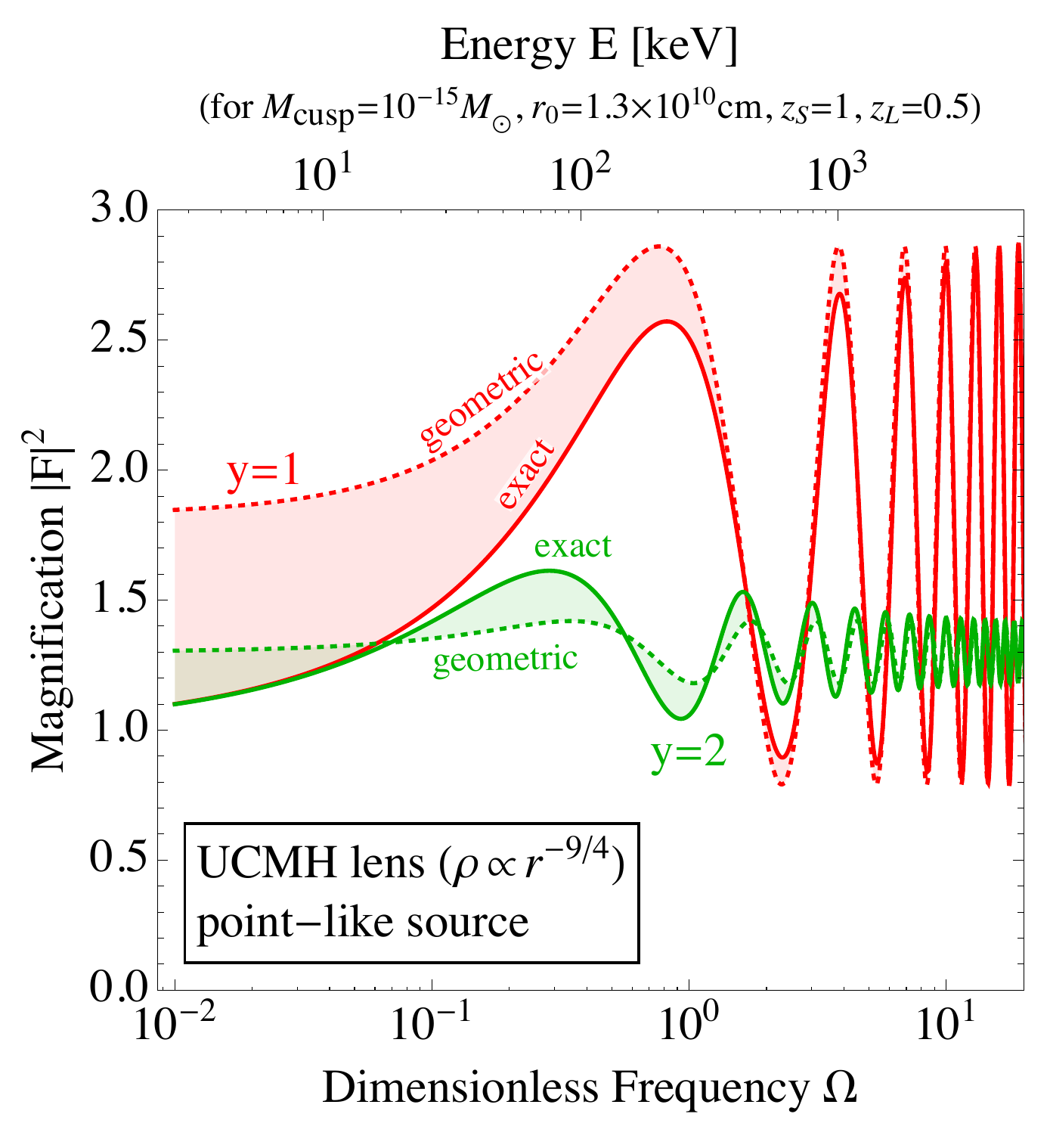}
  \caption{True magnification vs.\ geometric optics approximation for the
    self-similar infall profile $\rho(r) \sim r^{-9/4}$ as a function of the
    dimensionless frequency $\Omega=\omega \, \Delta t_0$, see
    \cref{eq:Omega_def_gen}.  We see that, for impact parameter
    $y\equiv\beta/\theta_0=1$, the picture is qualitatively quite similar to
    that for a point-like mass. At larger $y$, however, the geometric optics
    approximation significantly \emph{underestimates} the magnitude of the
    lensing effect in a wide frequency range.}
  \label{fig:94_phys_vs_geom}
\end{figure}

Wave optics effects quantitatively modify the geometric optics results at low
frequencies, but do not lead to qualitative changes. We have computed the
magnification numerically for the case $\delta=9/4$ using
\cref{eq:magnification-wave,eq:physopt}.  The results are shown in
\cref{fig:94_phys_vs_geom}, and are also compared to the geometric optics
approximation.  As expected from the above discussion, the behavior of the
curves is similar to the case of a point-like lens, but it nevertheless differs
in some important details.  First, wave optics corrections remain sizeable up
to higher energies.  Even more important, for impact parameters $y \equiv
\beta/\theta_0 > 1$, we find that the geometric optics approximation
significantly  \emph{underestimates} the amplitude over the first 5--7 periods
of the oscillation. As the impact parameter grows, the amplitude of the
oscillations falls off more slowly towards higher frequency than for a
point-like lens,  increasing the lensing probability for UCMHs compared to
PBHs.

Comparing \cref{fig:BH_phys_vs_geom,fig:94_phys_vs_geom}, we observe that the
interference fringes for UCMHs are shifted to higher energies compared to
the case of PBHs.  This
can be understood by noting from \cref{eq:lensing-potential} that a photon
passing the lens at a distance $\theta D_L$ is affected by DM
particles residing at radii less than $\theta D_L$.  The characteristic distance
at which lensed photons pass an UCMH is $\theta_0 D_L$ (see \cref{eq:theta-0-UCMH}).
So if we call the mass contained within this radius $m(\theta_0)$, we expect
the interference pattern for an UCMH to be comparable to the one for a PBH
with mass $m(\theta_0)$ (modulo obvious differences in the lensing potential 
and the reduced deflection angle).  We can thus estimate that the interference
fringes in \cref{fig:94_phys_vs_geom} ($\delta = 9/4$) should be shifted to higher
energies by a factor
\begin{align}
  \frac{m(\theta_0)}{M_\text{cusp}}
    = \bigg( \frac{\theta_0 D_L}{r_0} \bigg)^{3/4} \,.
  \label{eq:m-over-M-UCMH}
\end{align}
compared to the ones in \cref{fig:BH_phys_vs_geom}.  For $z_S = 1$ and $z_L =
0.5$, this becomes
\begin{align}
  \frac{m(\theta_0)}{M_\text{cusp}}
    = 0.037 \times \bigg( \frac{1.3 \times 10^{10}\,\text{cm}}{r_0} \bigg)^{6/5}
                   \bigg( \frac{M_\text{cusp}}{10^{-15} M_\odot} \bigg)^{3/5} \,.
  \label{eq:m-over-M-UCMH-numerical}
\end{align}
This estimate is indeed in good agreement with the shift observed in the plots.

\section{Revision of Current Bounds and Sensitivity Estimates}
\label{sec:bounds}

In this section, we will revisit the femtolensing bounds from
ref.~\cite{Barnacka:2012bm} by considering wave optics corrections (see
\cref{sec:wave-optics}) as well as the non-pointlike nature of the GRB sources
(see \cref{sec:extended-source}).  From Fig~\ref{fig:BH_phys_vs_geom} we expect
that wave optics effects will modify femtolensing bounds on point-like masses
by at most a few tens of per cent compared to the geometric optics
approximation.  The finite size of the sources, however, is expected to lead to
much more dramatic modifications.  As the size $a_S$ of the emission region in
a GRB is very uncertain (see \cref{app:GRBsize}),
we will investigate the dependence of our results on
$a_S$. We will find that only if $a_S \lesssim 10^8$~cm, current data is able
to set meaningful limits. This is true for point-like lenses such as primordial
black holes, but also for extended lenses like axion miniclusters.
Unfortunately the assumption $a_S \lesssim 10^8$~cm is not realistic.

We will establish these conclusions by investigating the \emph{sensitivity} of
current data, i.e. by working with simulated GRB data rather than real data.
This will make our analysis more transparent, and our conclusion will be that
current data is not sensitive to DM femtolensing yet. Therefore it is not
necessary to go beyond a sensitivity study, and we will therefore not analyze
actual Fermi GRBs data.  We will, however, extrapolate our results into the
future and estimate how many well observed GRBs would be needed for
femtolensing to become a competitive player in the hunt for compact DM objects.

In the following, we will first describe how we model GRB spectra 
(\cref{sec:grb-models}).  We will then describe the statistical methods
we use (\cref{sec:likelihood}) and discuss the resulting sensitivity estimates
for PBHs and UCMHs (\cref{sec:sensitivity-results}).

\subsection{GRB models}
\label{sec:grb-models}

We simulate ``data'' based on a phenomenological model for the unlensed GRB
spectrum. In particular, we use Band's model (BAND)~\cite{Band:1993eg} as our
baseline scenario, but we have also studied a broken power law model (BKN) as
well as a simple power law model with an exponential cutoff as a cross check. 

The BAND model has four free parameters: an amplitude
$A$, two spectral indices $\alpha_1$ (low energy) and $\alpha_2$ (high energy),
and an energy scale $E_0$. In terms of these parameters, the spectrum as a function
of energy $E$ is given by
\begin{align}
   f_\text{BAND}(E) &= 
     \begin{dcases}
       A \, (E/E_0)^{\alpha_1} \exp(-E/E_0)
                        & \text{if $E \leq (\alpha_1 - \alpha_2) E_0$} \\
       A \big[(\alpha_1 - \alpha_2) \big]^{\alpha_1-\alpha_2} 
         (E/E_0)^{\alpha_2} \exp(\alpha_2 - \alpha_1)
                        & \text{otherwise}
     \end{dcases} \,.
  \label{eq:band}
\end{align}
In our sensitivity studies, we choose $A = 0.15~\text{counts}\
\text{sec}^{-1}\,\text{cm}^{-2}\,\text{keV}^{-1}$, $E_0 = 160$~keV, $\alpha_1 =
-0.9$, and $\alpha_2 = -2.5$.  These parameters are based on a fit to Fermi GBM
data on GRB~090424~\cite{Bhat:2016odd}, and we have normalized the spectrum to
5\,000 photons in the energy range from 8~keV to 550~keV on which we will focus
in our analysis.  This normalization is roughly based on the sample of high quality events that the 
Fermi GBM can collect for a typical short  GRB at $z_S = 1$. 
We treat the detector's effective area as a constant,
$100~\text{cm}^2$, and we assume the GRB to last $1~\text{sec}$. These are
typical values for short GRBs, which may appear more interesting for
femtolensing because of the arguments given in \cref{app:GRBsize} that indicate
that the size of the emission region, $a_S$, tends to be smaller for short
GRBs.  Nonetheless, given the possibility of a broad distribution of $a_S$ for
long GRBs, their larger abundance and higher redshift, it is unclear which type
of GRB will ultimately offer the best sensitivity. 

The BKN model has four free parameters as well: an amplitude $A$, a
characteristic energy $E_0$, and two spectral indices $\alpha_1$,
$\alpha_2$:
\begin{eqnarray}
   f_\text{BKN}(E) &=  
   \begin{dcases} 
       A \left( \frac{  E}{ E_0} \right)^{ -\alpha_1} & \text{if } E \leq E_0 \\
       A \left( \frac{  E}{ E_0} \right)^{ -\alpha_2} & \text{if } E  >   E_0 
   \end{dcases} \,.
\end{eqnarray}
The benchmark parameters for the BKN model are
$A = 0.099~\text{counts}\
\text{sec}^{-1}\,\text{cm}^{-2}\,\text{keV}^{-1}$,
$E_0 = 160$~keV, $\alpha_1 = -0.9$, and $\alpha_2 = -2.5$.

The power law model with exponential cutoff has only three
free parameters: the amplitude $A$, a spectral index $\alpha$, and an energy scale
$E_0$:
\begin{align}
  f_\text{power-exp} &= A \, E^{\alpha} \exp(-E/E_0) \,.
  \label{eq:power-exp}
\end{align}
The benchmark parameters for this model are $A = 0.16~\text{counts}\
\text{sec}^{-1}\,\text{cm}^{-2}\,\text{keV}^{-1}$, $E_0 = 150$~keV, and
$\alpha = -0.92$. This choice leads to the same normalization as for the BAND
model.\footnote{The power law model is not strongly motivated physically, and is
  just used here as a cross check to verify the robustness of our predictions.
  In particular,
  if there is an exponential cutoff in the GRB spectrum, it is expected at an energy
  much higher than $150$~keV~\cite{Nava:2018qkq}.}

We choose energy-dependent bin sizes of $2 \, \delta E$, where we assume
\begin{align}  \label{eq:resolution}
  \frac{\delta E}{E} = \sqrt{ \bigg( \frac{0.05}{\sqrt{E / \text{keV}}} \bigg)^2
                                   + (0.05)^2 } \,.
\end{align}
This resolution is better than that of the Fermi GBM
instrument~\cite{Meegan:2009qu} because we will mainly be interested in the
sensitivity of future observatories.

\begin{figure}
  \centering
  \includegraphics[width=0.6\columnwidth]{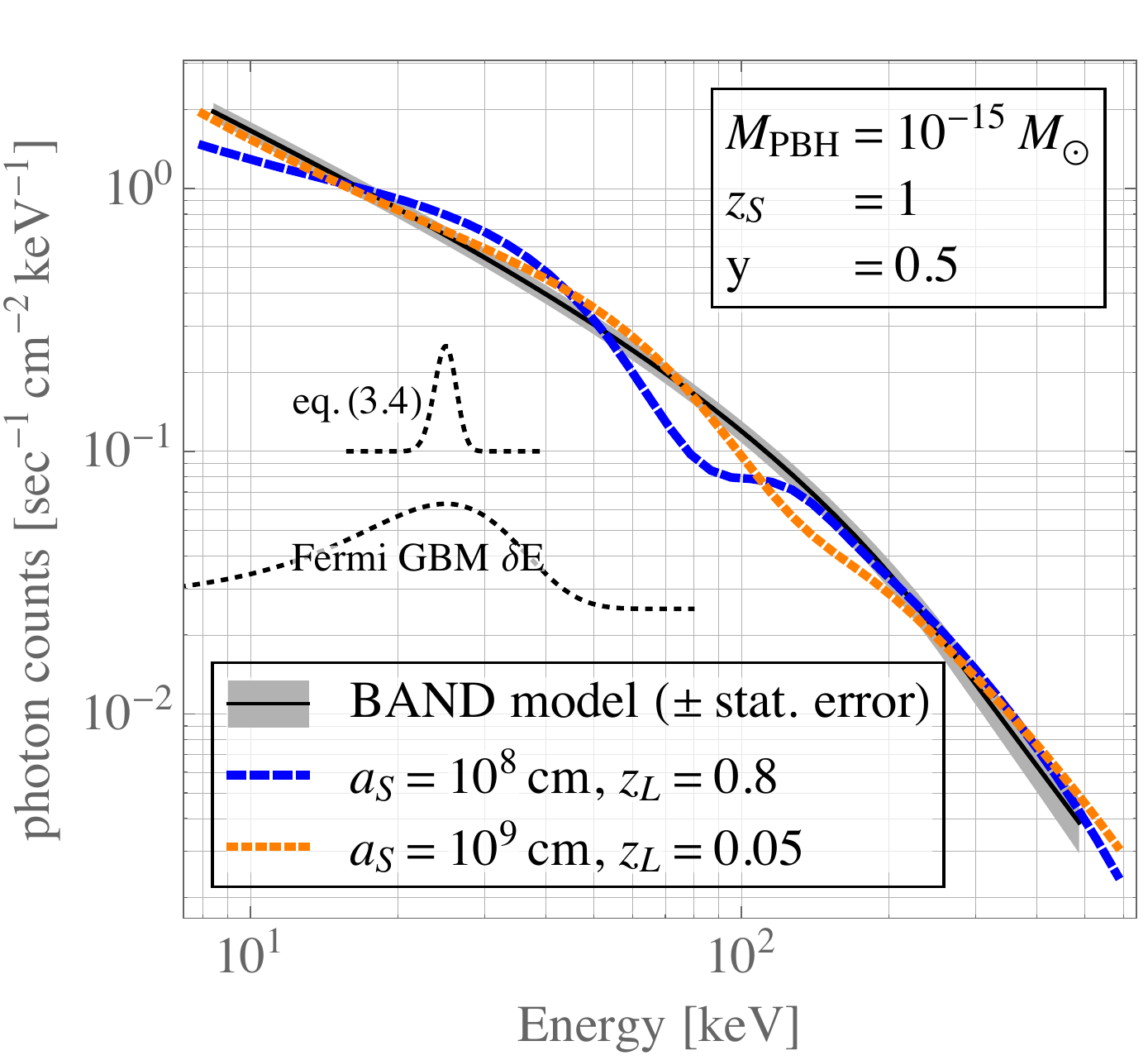}
 \caption{Simulated GRB spectra based on the BAND model with parameters
    $A= 14.08~\text{counts}\ \text{sec}^{-1}\,\text{cm}^{-2}\,\text{keV}^{-1}$,
    $E_0 = 160$~keV, $\alpha_1 = -0.9$, and $\alpha_2 = -2.5$. We compare the unlensed
    spectrum (black) to the predicted spectra in the presence of a PBH lens
    with mass $M = 10^{-15}~M_\odot$ and impact parameter $y = 0.5$.
    Wave optics effects as well as the finite size of the GRB emission region
    are taken into account, where for the latter we use either $a_S = 10^8$~cm
    (blue) or $10^9$~cm (orange). The source is assumed to be
    located at redshift $z_S = 1$. For highly optimistic (small) $a_S$, a pronounced    interference pattern is visible almost independently of the position of the
  lens, while for more realistic (larger) $a_S$, only a lens rather close
    to the observer may lead to an observable effect.
    The gray band drawn around the unlensed data represents the statistical
  uncertainty.The black dotted gaussians show the detector response assuming 
the FERMI energy resolution as well as the resolution of eq.~\eqref{eq:resolution} that we 
optimistically assume for future detectors. The signal that one observes in the detector would be a 
convolution of  the detector response with the line simulated GRB spectrum.}
  \label{fig:databs}
\end{figure}

In \cref{fig:databs}, we show a simulated data set based on these assumptions
and using the BAND model. We see that the magnitude of the femtolensing effect
depends crucially on the size of the source.  If $a_S = 10^8$~cm, a pronounced
interference pattern emerges. However already for $a_S \sim 10^9$~cm -- which
is still rather optimistic -- it is already quite difficult to extract a clear
signal, even if the lens is located at very low redshift $z_L = 0.05$, so that
the projected size of the source in the lens plane is reduced compared to the
case $z_L \sim 1$.  Had we chosen the same $z_L$ for the orange curve as for
the blue one ($a_S \sim 10^8$~cm), the amplitude of the femtolensing wiggles
would be comparable to the error bars on the data or even smaller.  Similarly,
for larger $a_S \sim 10^{10}$~cm, the femtolensing effect disappears.  Parenthetically
we  notice that in order to get the realistic pattern, one should further convolve it 
with the detector response function (see the black dotted lines). We checked explicitly, that 
the resolution that we have assumed in eq.~\eqref{eq:resolution} does not change qualitatively 
any of the curves that we show, the current Fermi resolution wipes out the effect 
completely for both lines. This further reinforces our point that
current Fermi observation cannot   
even be sensitive to the emission size as small as $a_s = 10^8$~cm
(which is extremely optimistic),  
and better resolutions of 
future experiments would be needed to make any further progress.

\subsection{Likelihood Analysis}
\label{sec:likelihood}

To quantitatively analyze the simulated data and compare to femtolensing
predictions, we define a log-likelihood function\footnote{Note that our
statistical procedure differs from the one in \cite{Barnacka:2012bm}. Our approach
combines the likelihood of the lensed signal for every possible 3-dimensional position
of the lens, weighted
by the probability of finding a lens at a given position. Ref.~\cite{Barnacka:2012bm},
in contrast, investigates the observability of lensing as a function of the lens mass
and the transverse position of the source, but without fully taking into account the
dependence on the redshift of the lens $z_L$.}
\begin{align}
  -2 \log L_0(\vec{\mu}_s) \equiv
    \min_{\vec\mu_b} \bigg[
      \sum_{j=1}^{\text{\# of bins}} \bigg(
                                       \frac{O_j - P_j(\vec{\mu}_b, \vec{\mu}_s)}{\sigma_j}
                                     \bigg)^2
    \bigg]  +  \text{const} \,,
  \label{eq:likelihood-1}
\end{align}
where $O_j$ (``number of observed events'') denotes the number of events in the
$j$-th bin without lensing, and $P_j(\vec{\mu}_b, \vec{\mu}_s)$
(``number of predicted events'') denotes the number of events expected if there
is a lens. $\sigma_j$ is the uncertainty associated with the $j$-th
bin, which is given by adding in quadrature the statistical uncertainty
$\sqrt{P_j(\vec{\mu}_b, \vec{\mu}_s)}$ and a systematic uncertainty. The latter
accounts for quasi-random deviations of the source spectrum from the model,
and we assume its magnitude to be 5\%. (We will also show how our results
change if the systematic uncertainty is set to 0\% or 10\% instead.)
The vector $\vec\mu_b$ contains the relevant parameters of the background model
(four for the BAND and BKN models, three for the power law model with exponential
cutoff).  To be conservative, we
minimize over $\vec\mu_b$, i.e.\ we choose the background parameters that best
fit the data.  The vector $\vec\mu_s$ contains the parameters of the lens, namely
its mass $M$, its redshift $z_L$, and its normalized impact parameter in the lens plane
$y = \beta / \theta_E$. Thus, $L_0$ compares the unlensed spectrum to the
lensed spectrum for fixed lens and source parameters.  Note that we can use
Gaussian rather than Poisson statistics here because the number of photon events
per bin is large.

We can define a lensing cross section $\sigma(D_L) = \pi (y_\text{max} \theta_E
D_L)^2$, where $y_\text{max}$ is the maximal normalized distance from the
lens to observer--source line of sight that still leads to
a sizable lensing signal. (A sizable signal is defined as a signal
that can be distinguished, at a given confidence level (CL) from an unlensed signal.)
In other words, $y_\text{max}$ is obtained by solving
\begin{align}
  -2 \log\bigg( \frac{L_0(\vec\mu_s | y = y_\text{max})}
                     {L_0(\vec\mu_s | y = \infty)} \bigg) = \alpha \,,
  \label{eq:lensing-xsec}
\end{align}
where $\alpha$ is the quantile one of the $\chi^2$ distribution
with one degree of freedom corresponding to the chosen CL. For instance,
$\alpha = 2.7$ for 90\% CL and $\alpha = 9$ for $3\sigma$ CL.
The optical depth $\tau$ of the source is obtained as
\begin{align}
  \tau = \int_0^{z_S} \! \frac{dz_L}{H(z_L)} \, \sigma(D_L)
                         \frac{\rho_\text{PBH}}{M} (1 + z_L)^2 \,,
  \label{eq:optical-depth}
\end{align}
where $\rho_\text{PBH} / M$ is the number density of lenses at the
present epoch. We integrate over the redshift of the lens, which is equivalent,
up to the Hubble constant $H(z_L)$, to integrating over comoving distance.
The factor $(1+z_L)^2$ appears as a combination of a 
factor $(1+z_L)^3$ that accounts for the increase in lens density
with redshift and a factor $(1+z_L)^{-1}$ needed to convert between
the physical and comoving longitudinal coordinate.

To obtain limits on the density of lenses, we need to take into account the
dependence of $L_0$ on $\vec\mu_s$.  To this end, we define the overall likelihood
for lensing of a single GRB source with size $a_S$ and redshift $z_S$ according to
\begin{align}  
  L^\text{1-GRB}(M, \rho_\text{PBH}, a_S, z_S) &\equiv L_0(0) +
    \int \! d^3x \, \frac{\rho_\text{PBH}}{M} (1+z_L)^3
      \big[ L_0(\vec\mu_s) - L_0(0) \big] \,.
  \label{eq:llpbh}
\end{align}  
Here, the integral runs over physical (not comoving) coordinates, and the
factor $(1+z_L)^3$ once again takes into account the increase of the lens
density with redshift.  The integration region extends from the observer to the
source in the longitudinal direction, and out to infinity in the transverse
directions.  We here make the (realistic) assumption that the probability of a
single GRB being lensed by multiple compact DM objects is $\ll 1$.

Of course, the sensitivity can be significantly boosted by observing not a
single GRB, but many of them. For $n$ observed GRBs, the total likelihood is
\begin{align}
  L^\text{$n$-GRBs}(M, \rho_\text{PBH}, \{a_{S,k} \}, \{ z_{S,k} \}) &\equiv
    \prod_{k=1}^{n} L^\text{1-GRB}(M, \rho_\text{PBH}, h_{S,k}, z_{S,k}) \,,
  \label{eq:L-nGRBs}
\end{align}
where $a_{S,k}$ and $z_{S,k}$ denote the size and redshift of the $k$-th GRB.

We set 95\% CL limits on the mass and density of the compact DM objects
by equating the log-likelihood ratio with the 95\% quantile
of the $\chi^2$ distribution with two degrees of freedom ($M$ and $\rho_\text{PBH}$):
\begin{align}
  -2 \log\bigg( \frac{L^\text{$n$-GRBs}(M, \rho_\text{PBH}, \{ a_{S,k} \}, \{ z_{S,k} \})}
                     {L^\text{$n$-GRBs}(0, 0, \{ a_{S,k} \}, \{ z_{S,k} \})}
         \bigg) &= 5.99
  & \text{(95\% CL)} \,.
  \label{eq:llr}
\end{align}
Here, the denominator corresponds to the likelihood of the data in the absence
of any lensing.

\subsection{Results}
\label{sec:sensitivity-results}

\begin{figure}
  \centering
  \includegraphics[width=0.6\columnwidth]{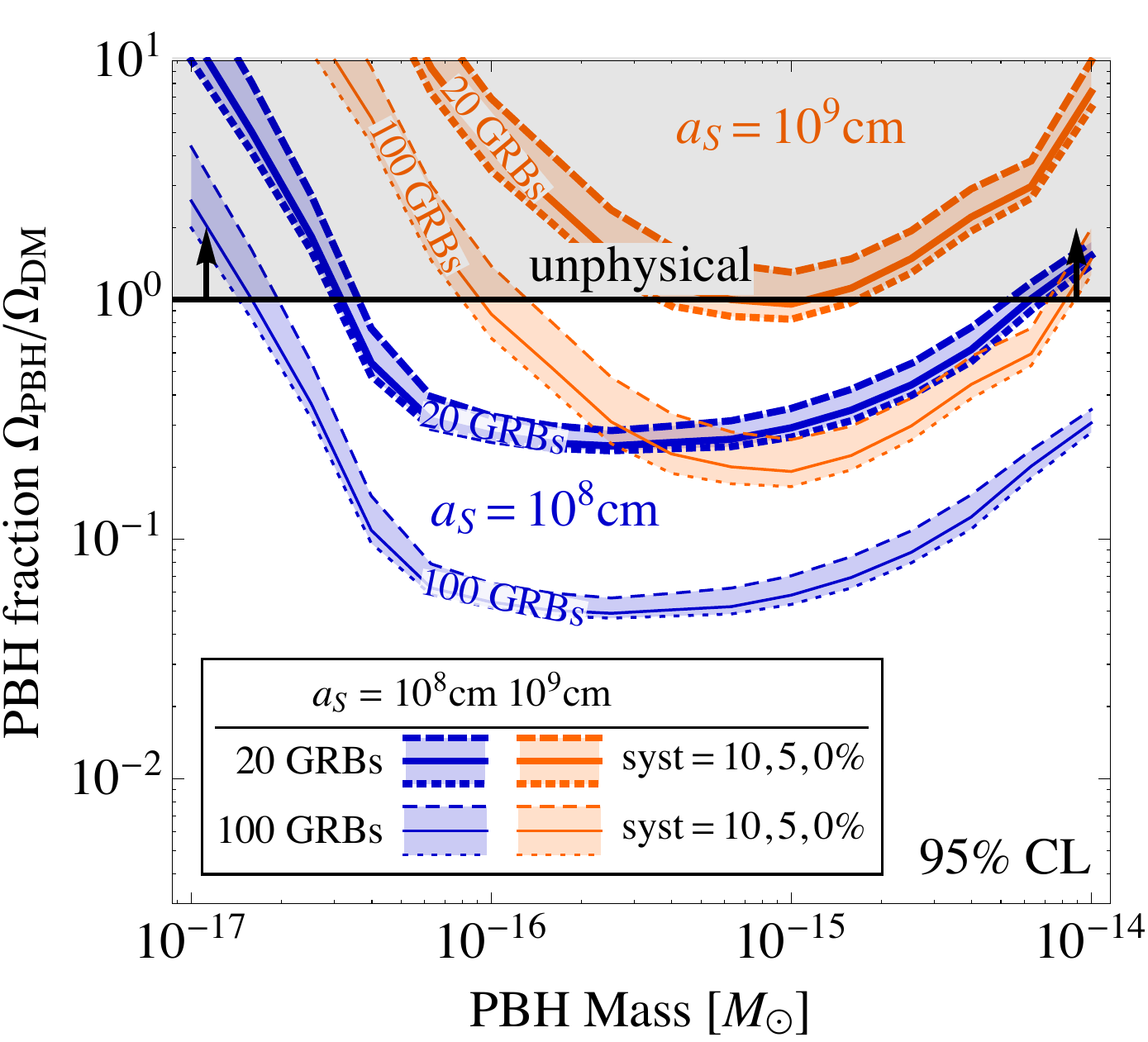}
  \caption{Sensitivity of femtolensing searches to the primordial black
    hole contribution $\Omega_\text{PBH} / \Omega_\text{DM}$ to the overall DM
    density in the Universe.  We show the projected sensitivity for different
    assumptions on the number of suitable GRBs with well-measured redshifts in
    the data sample.  We also illustrate the dependence on the size $a_S$ of
    the emission region of a typical GRB, where $a_S = 10^8$~cm should be
    considered a highly optimistic value, and $a_S = 10^9$~cm an optimistic but
    possible value.  The colored bands indicate the impact of systematic
    effects (uncorrelated random fluctuations in each bin).  We use the BAND
    model for the GRB spectrum throughout.  For other spectral models, limits
    would change by not more than a few tens of percent.  We have assumed the
    redshift of all GRBs in the sample to be $z_S = 1$.}
  \label{fig:contour1}
\end{figure}

With the above likelihood formalism in hand, we can now study the sensitivity
of current and future data to compact DM objects.  We focus on the mass range
$10^{-17} M_\odot$ to $10^{-14} M_\odot$.  This mass window is motivated by the
requirement that a sizable phase difference between the different lensed
images of the source should occur within the energy range of GRB spectra.  Our
results are shown in \cref{fig:contour1}. We see that with 20~GRBs with
well-measured redshifts (the number of GRBs used in \cite{Barnacka:2012bm}), a
meaningful limit can only be set if the GRBs are assumed to be point-like,
i.e.\ $a_S \lesssim 10^8$~cm.  Here, by ``meaningful limit'' we mean a limit
that constrains the cosmological PBH abundance, $\Omega_\text{PBH}$, to be less
than the total DM abundance in the Universe, $\Omega_\text{DM}$.  As argued in
\cref{sec:extended-source}, the assumption $a_S \lesssim 10^8$~cm is almost
certainly overly optimistic.  If $a_S$ is only one order of magnitude larger
(which is still very optimistic), no limit can be set.

This may change in the future if the available sample of GRBs is
significantly extended. With 100 GRBs, sensitivity to an $\mathcal{O}(30\%)$
fraction of PBHs can be achieved for $a_S \sim 10^9$~cm.  The sensitivity
would improve to $\Omega_\text{PBH} / \Omega_\text{DM} \sim 0.06$ if $a_S
\sim 10^8$~cm.  Our conclusions are essentially independent of the choice of
GRB model (BAND vs.\ BKN vs.\ power law with exponential cutoff). They depend somewhat on the
assumed systematic uncertainty, with the sensitivity deteriorating by at
most a factor of two if the assumed systematic error is increased from 0\% to
10\%.  Note that we have very conservatively assumed systematic errors to be
completely uncorrelated between energy bins.

\begin{figure*}
  \hspace*{-2.5cm}    
  \includegraphics[width=1.3\columnwidth]{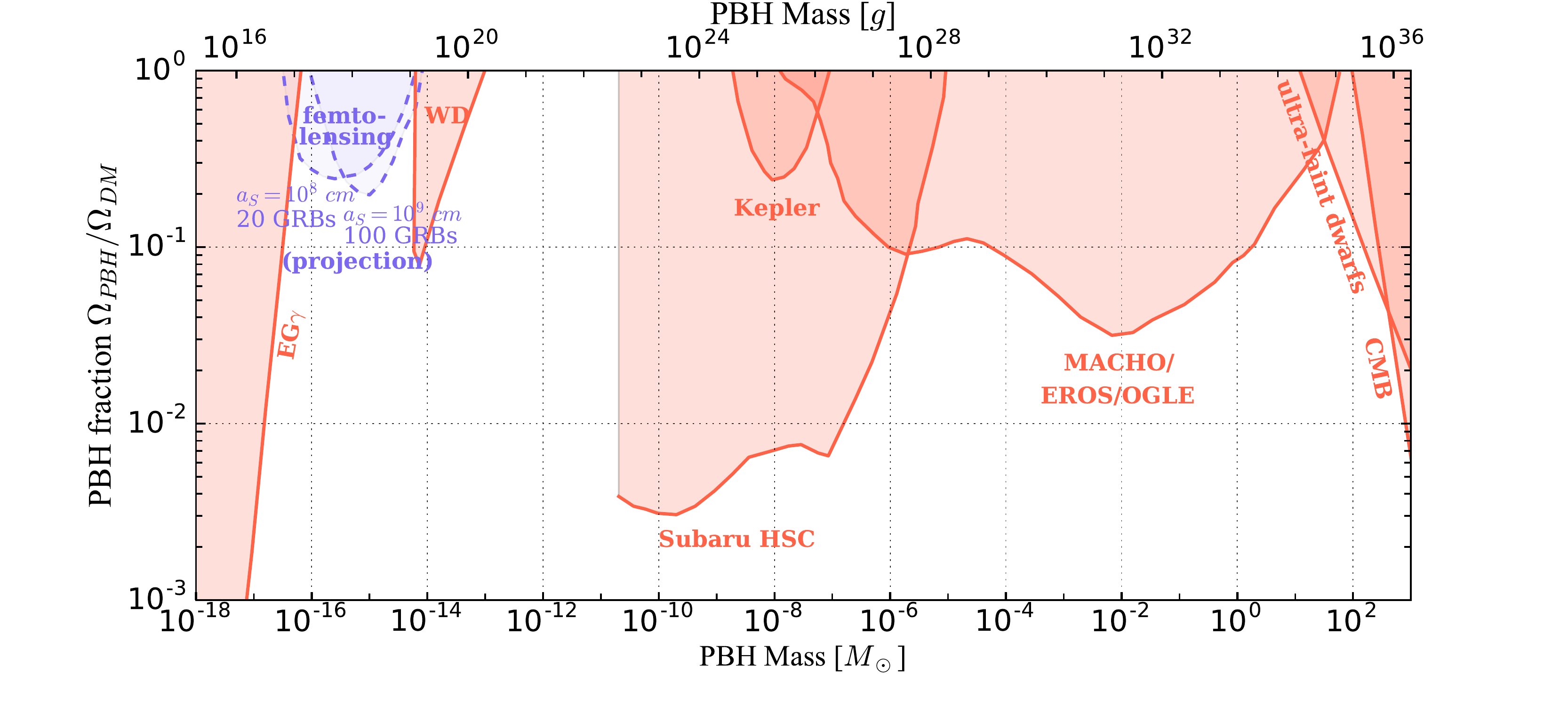}
  \caption{Future femtolensing sensitivity to primordial black holes compared
    to other probes.  In particular, we compare our projected limits (blue
    dashed contours) to limits based on extragalactic background photons
    (EG$\gamma$) from PBH evaporation \cite{Carr:2009jm}, from the
    non-destruction of white dwarfs (WD) \cite{Graham:2015apa}, from
    microlensing searches by Subaru HSC \cite{Niikura:2017zjd}, Kepler
    \cite{Griest:2013esa}, MACHO \cite{Allsman:2000kg}, EROS
    \cite{Tisserand:2006zx}, and OGLE~\cite{Wyrzykowski:2011tr}, from the
    dynamics of ultra-faint dwarf galaxies~\cite{Brandt:2016aco}, and from CMB
    anisotropies 
due to accretion onto PBHs~\cite{Ali-Haimoud:2016mbv}.  (Stronger
    CMB limits are obtained if more aggressive assumptions on accretion by
    PBHs are adopted~\cite{Poulin:2017bwe}.)  The
    Subaru HSC limits are cut off at $M \sim 10^{-11} M_\odot$ because below
    that mass, the geometric optics approximation employed in
    ref.~\cite{Niikura:2017zjd} is not valid.
    We also do not include neutron star limits \cite{Capela:2013yf} because of their
    dependence on controversial assumptions about the DM density in globular clusters.
    We have taken the
    limits shown here from
    the compilation in ref.~\cite{Inomata:2017vxo}.  In computing our projected limits,
    we have assumed the redshift of all GRBs in the sample to be $z_S = 1$, we
    have used the BAND model for the GRB spectrum, and we have assumed a 5\%
    systematic uncertainty, uncorrelated between energy bins.}
  \label{fig:pbh-all-constraints}
\end{figure*}

In \cref{fig:pbh-all-constraints}, we put our projected constraints into a
broader context by comparing to other limits on PBHs.  We see that future
femtolensing constraints, albeit weak, may cover a mass range that is otherwise
inaccessible and where viable PBH DM could exist.

\vspace{2ex}
While the projected limits shown in \cref{fig:contour1,fig:pbh-all-constraints}
apply only to PBH DM, we can use the arguments given at the end of
\cref{sec:extended-lens} as a starting point for estimating also the
sensitivity to other compact DM structures.  For UCMHs, we have argued above
that the interference fringes are shifted to higher energies compared to the
PBH case, with the magnitude of the shift, $m(\theta_0)/M_\text{cusp}$, given
by \cref{eq:m-over-M-UCMH,eq:m-over-M-UCMH-numerical} for UCMHs with $\rho(r)
\propto (r_0/r)^{9/4}$. We therefore estimate that the projected femtolensing
exclusion limits on such UCMHs will be similar in shape to the ones for PBHs if
the PBH mass is understood as the ``equivalent black hole mass'' $m(\theta_0)$
from \cref{eq:m-over-M-UCMH,eq:m-over-M-UCMH-numerical}. Expressed in terms of
$M_\text{cusp}$, the limits on UCMHs are thus shifted to higher masses by a
factor $M_\text{cusp} / m(\theta_0)$ compared to the limits shown in
\cref{fig:contour1,fig:pbh-all-constraints}.  We moreover need to take into
account the fact that the DM number density scales inversely with mass, hence
we expect the projected exclusion curves to also move upwards by a factor
$M_\text{cusp} / m(\theta_0)$, corresponding to a weakening of the limit by
that factor.  In view of this, and given that we have seen how difficult it
will already be to constrain PBH DM using femtolensing, we refrain from a more
detailed sensitivity study for UCMHs.

\section{Conclusions}
\label{sec:conclusions}

To summarize, we have critically investigated the potential of gravitational
femtolensing to constrain compact DM structures such as primordial black holes
or ultracompact DM minihalos.  Femtolensing exploits the tiny time delay
between the multiple lensed images of a distant source. Interference between
the images leads to characteristic fringes in the energy spectrum of observed
photons, see \cref{fig:databs}.  As sources, we consider in particular gamma
ray bursts at $\mathcal{O}(1)$ redshift.  These sources are most easily
observed at energies $\gtrsim 10$~keV, hence observable interference pattern
require time delays $\lesssim 10^{-19}$~sec.  This means that the best
sensitivity is expected for compact DM objects in the mass range from
$10^{-17}$ to $10^{-14}~M_\odot$.

We have argued that the simple geometric picture of femtolensing based on
point-like sources and lenses that is often used in the literature is not
appropriate in reality.  First, it is not true that photons travel from the
source to the detector along one of just two discrete paths.  In fact, when the
time delay becomes comparable to the inverse photon frequency (which for
point-like lenses is equivalent to the photon wave length becoming comparable
to the Schwarzschild radius of the lens), wave optics effects become
non-negligible. It is then necessary to integrate the photon amplitude over the
whole lens plane. This leads to $\mathcal{O}(1)$ corrections to the
interference pattern at the lower end of the photon energy spectrum.  Second,
while the approximation of a point-like lens works for primordial black holes,
it is not satisfied for ultra-compact mini-halos, and even less so for NFW-like
structures.  We have therefore computed femtolensing effects for generic
power-law density profiles, and have explicitly shown numerical results for the
self-similar infall profile with $\rho(r) \propto r^{-9/4}$.

The most important correction in femtolensing of GRBs is coming from
the non-negligible size $a_S$ of the GRB source itself. In fact, we have argued
that a GRB could only be treated as point-like for the purpose of femtolensing
if the photon emission region was smaller than $a_S \sim 10^8$~cm. And while
estimates for the size of the emission region can vary by a few orders of
magnitude, $a_S \sim 10^8$~cm seems unrealistically small. For more realistic
assumptions on the value of $a_S \gtrsim 10^{10}$~cm, the femtolensing effect
is almost entirely washed out. This means that, contrary to previous claims,
current GRB data is insufficient to constrain compact DM structures such as
primordial black holes, even if their abundance $\Omega_\text{PBH}$ saturates
the observed DM abundance $\Omega_\text{DM}$ in the Universe.  We have,
however, demonstrated that constraints down to $\Omega_\text{PBH} /
\Omega_\text{DM} \sim 0.2$ would become possible with a sample of about 100
observed GRBs with well-measured redshifts and spectra, and with small $a_S$.
Such GRBs are expected to be characterized by very fast intrinsic
variability at sub-millisecond time scales.

Since femtolensing constraints on compact DM objects are still out of reach we
conclude that there are currently \emph{no} firm bounds on PBH DM in the mass
range of $10^{-17} \div 10^{-11}$ (with the exception of a small wedge of
parameter space excluded by white dwarf observation). This is illustrated in
\cref{fig:pbh-all-constraints}.  While several ideas has been put forward to
constrain this mass range, for instance picolensing \cite{Nemiroff:1995ak} and
the capture of PBHs by stars or neutron stars, none of these methods has
yielded decisive bounds until now, so this region of parameter space still
awaits exploration.

\acknowledgments

We are grateful to Yacine Ali-Ha\"imoud, Juan Garcia Bellido, Kfir
Blum, Lam Hui and
Kathryn Zurek for useful discussions. JK has been supported by the German
Research Foundation (DFG) under Grant  Grant Nos. KO 4820/11, FOR 2239,
EXC-1098 (PRISMA) and by the European Research Council (ERC) under the European
Union's Horizon 2020 research and innovation programme (grant agreement No.
637506, ``$\nu$Directions'').  JK would also like to thank CERN for hospitality
and support.

\appendix
\section{The size of the prompt emission region in GRBs}
\label{app:GRBsize}

In this appendix we discuss the transverse size of the $\gamma$-ray emitting
regions in GRBs, which, as we have seen in \cref{sec:extended-source}, plays a
crucial role in the study of femtolensing. We will review estimates for the
size of the emission region based on measurements of the variability time
scale, and we will also discuss the lower bound following from the required
transparency of the emission region to $\gamma$-rays.

It is believed that the prompt $\gamma$-emission of GRBs is produced by
electrons and positrons accelerated in relativistic shock waves. The
non-thermal $\gamma$-ray spectrum implies that the emission region must be
optically thin.  To reconcile this requirement with the observed energetics of
the bursts, the bulk Lorentz factor of the $\gamma$-ray emitting material must
be large, $\varGamma\gtrsim 100$~\cite{Piran:2004ba}. 

We etsimate the transverse size of the emission region following the
same approach as Ref.~\cite{Barnacka:2014yja}.
Consider a blob of material of size $a_S$ moving with velocity $v$ at an angle
$\theta_\text{obs}$ to the observer's line of sight. In the rest frame of the
blob the minimal variability time scale of the emission is simply given by the
light-crossing time, 
\begin{align}
  \hat t_\text{var}\sim a_S /c \,.
  \label{eq:tvarhat}
\end{align}
The observed variability time $t_\text{var}$ is related to  $\hat t_\text{var}$
by the relativistic Doppler formula,
\begin{align}
  t_\text{var} = (1+z_S) \bigg(1 - \frac{v}{c} \cos\theta_\text{obs}\bigg)
                 \varGamma \, \hat t_\text{var}\,,
  \label{eq:tvar}
\end{align}
where $\varGamma \equiv \big(1 - (v/c)^2\big)^{-1/2} $ and we have taken into
account the cosmological redshift of the source. Due to the relativistic
beaming effect, we have $\theta_\text{obs}\sim 1/\varGamma$. Combining this
with \cref{eq:tvarhat,eq:tvar} one obtains the estimate
\begin{align}
  a_S \sim \frac{c \varGamma \, t_\text{var}}{1+z_S}
      \simeq \frac{10^{11} \, \text{cm}}{1+z_S} \times
             \bigg( \frac{t_\text{var}}{0.03\,\text{sec}}\bigg)
             \bigg( \frac{\varGamma}{100}\bigg) \,.
  \label{eq:ahatestim}
\end{align}

The minimal variability time scales for various GRBs have been determined in
ref.~\cite{Golkhou:2015lsa}. They lie within the range
$t_\text{var}^\text{sGRB} \sim (0.01\div 0.1)\;\text{sec}$ for short GRBs and
$t_\text{var}^\text{lGRB} \sim (0.1\div 1)\;\text{sec}$ for the long ones.
These results are consistent with the earlier estimates of
ref.~\cite{Barnacka:2014yja} that give average variability time scales
$t_\text{var}$ of $0.036$~sec and $1.2$~sec for short and long GRBs,
respectively. We see that for a typical short GRB with $z\sim 1$,
$t_\text{var}\sim 0.03$~sec, and $\varGamma \sim \mathcal{O}(100)$, the
transverse size is,
\begin{align}
  a_S \sim 10^{11} \; \text{cm} \,.
  \label{eq:ahatnum}
\end{align} 
For long GRBs this estimate becomes an order of magnitude larger.\footnote{Note
  that Ref.~\cite{Golkhou:2015lsa} derives also the distribution of the
  emission radii $R_\text{em}$, i.e.\ the distance from the GRB central engine
  at which $\gamma$-rays are emitted. The central value of this distribution is
  $R_\text{em}\sim 3\times 10^{13}$~cm ($10^{14}$~cm) for short (long) GRBs.
  Due to relativistic beaming, the transverse size of the patch visible by an
  observer on Earth is related to this distance as $a_S \sim
  R_\text{em}/\varGamma$. Assuming $\varGamma\sim {\cal O}(100)$, this again
  gives $ a_S\sim 10^{11}$\;cm ($10^{12}$\;cm) for short (long) GRBs.}

In the above estimates we adopted the standard picture of a GRB with a
relativistic jet of $\gamma$-ray emitting material pointing towards the
observer. One may wonder how the reasoning is modified in the case of off-axis
observation, like in the recent GRB 170817A
\cite{Goldstein:2017mmi,Savchenko:2017ffs}  accompanied by the gravitational
wave event GW170817 \cite{TheLIGOScientific:2017qsa}, which has confirmed the
identification of neutron star mergers as progenitors of short GRBs. As
discussed in \cite{Kasliwal:2017ngb}, the observational data pertaining to this
event across the electromagnetic spectrum are best described by a model in
which a mildly relativistic wide-angle shock with $\varGamma\sim 2.5$ breaks
off the ambient material and emits $\gamma$-rays at a distance
$R_\text{em}\simeq 2.4\times 10^{11}$\;cm from the central engine. In this case
relativistic beaming is practically absent, so $a_S$ is comparable to
$R_\text{em}$. We again recover the estimate (\ref{eq:ahatnum}).

If all GRBs satisfy the estimate~\eqref{eq:ahatnum}, observing any femtolensing
in their spectra will be essentially hopeless. So let us ask how robust this
estimate is. A hint that there may be GRBs with smaller sizes is already
provided by the analysis of Ref.~\cite{Golkhou:2015lsa}, which finds that about
10\%   of the bursts (both short and long) exhibit faster variability,
$t_\text{var} \lesssim 2\times 10^{-3}~\text{sec}$, which leads to an estimated
size $a_S \lesssim 10^{10}$~cm. Thus, one can speculate that some rare GRBs at
the tail of the distribution could have $t_\text{var}$ yet another order of
magnitude shorter, bringing their sizes close to the values of $a_S$ required
for efficient femtolensing.  It is also worth pointing out that the
determination of the minimal variability time scale is affected by instrumental
systematics, such as the detector sensitivity and the light-curve sampling.
Thus, it might happen, in principle, that some of the measured variability time
scales overestimate the true intrinsic variability time scale of the source. 

An alternative method for constraining $a_S$ is based on the requirement that
the emission region must be optically thin. Two processes can lead to
absorption of $\gamma$-rays: production of $e^+e^-$ pairs in
$\gamma\gamma$-scattering, and Compton scattering on electrons and positrons.
Let us start with the first process.

Consider a photon with energy $\hat E$ in the rest frame of the
emitting blob. The optical depth for pair production is,
\begin{align}
  \tau_{\gamma\gamma}(\hat E)
    = a_S \int_{\hat\epsilon_\text{th}} \! d\hat\epsilon \;
          \sigma_{\gamma\gamma}(\hat \epsilon) \, \hat n_\gamma(\hat\epsilon) \,,
  \label{eq:taugg}
\end{align}   
where $\sigma_{\gamma\gamma}(\hat\epsilon)$ is the (angular-averaged) cross
section of collision with an ambient photon of energy $\hat\epsilon$, 
$\hat n_\gamma(\hat\epsilon)$ is the spectral density of such photons,
and 
\begin{align}
  \hat\epsilon_\text{th} = \frac{m_e^2c^4}{\hat E}
\end{align}
is the threshold energy for pair production. To estimate $\hat n_\gamma(\hat
E)$ we notice that the number of photons with energy $\hat E$ emitted by the
blob during its proper time $d\hat t$ is
\begin{align}
  dN \simeq c \, a_S^2 \, \hat n_\gamma(\hat E) \, d\hat E \, d\hat t \,.
\end{align}
This is related to the photon flux $f(E)$ seen by the observer at energy
$E = \varGamma \hat E / (1+z_S)$ as
\begin{align}
  dN \simeq \frac{d_S^2}{\varGamma^2(1+z_S)^2} \, f(E) \, dE \, dt \,,
\end{align}
where $d_S$ is the luminosity distance of the GRB and the factor
$1/\varGamma^2$ is due to the fact that in the observer's frame the
emission is beamed into a narrow cone with opening angle $\sim
\varGamma^{-1}$. Next, due to the Doppler effect we have
(cf.\ \cref{eq:tvar}) $dt \simeq (1+z_S) \, d\hat t \, \varGamma$. Inserting all
these relations into (\ref{eq:taugg}) we obtain,
\begin{align}
  \tau_{\gamma\gamma}(\hat E)\simeq\frac{d_S^2}{c \, a_S\varGamma^2(1+z_S)^2}
  \int_{\hat\epsilon_\text{th}}d\hat\epsilon\;\sigma_{\gamma\gamma}(\hat\epsilon)\,
  f\big(\varGamma\hat\epsilon/(1+z_S)\big)\;.
  \label{eq:taugg1}
\end{align}
Following common practice, we now assume that the spectrum of the
prompt emission is described by a power law,
\begin{align}
  f(E) = A \, E^{\alpha} \,.
\end{align}
with spectral index $\alpha$ close to $-2$. Then the integral in
\cref{eq:taugg1} can be evaluated with the result,
\begin{align}
  \tau_{\gamma\gamma}(\hat E)\simeq\frac{d_S^2\sigma_T}{c
    a_S}\eta(\alpha)A\varGamma^{\alpha-2}
    (1+z_S)^{-\alpha-2}\bigg(\frac{m_e^2c^4}{\hat E}\bigg)^{\alpha+1}\;,
  \label{eq:taugg2}
\end{align}
where $\sigma_T$ is the Thomson cross section and the numerical
coefficient is \cite{Gould:1967zzb,1987MNRAS.227..403S}
\begin{align}
  \eta(\alpha) = \frac{3\big(\alpha(\alpha-5)(\alpha-1)-2\big)\sqrt{\pi} \, \Gamma(-\alpha)} 
                      {8\alpha(1-\alpha)(\alpha-2)(\alpha-1)\Gamma(3/2-\alpha)}\,.
\end{align}
Expressing the optical depth in terms of the photon energy measured by
the observer we arrive at,
\begin{align}
  \tau_{\gamma\gamma}(E) \simeq \frac{d_S^2\sigma_T}{c a_S} \eta(\alpha)
    \varGamma^{2\alpha-1} (1+z_S)^{-2\alpha-3} \,
    \frac{m_e^2c^4}{E} f\bigg(\frac{m_e^2c^4}{E}\bigg) \,,
  \label{eq:taugg3}
\end{align}
Up to a numerical coefficient of order one, this coincides with
the expression obtained in \cite{2001ApJ...555..540L}.
For $\alpha<-1$ the optical depth grows with the energy of the
photon. Requiring that it is smaller than one for photons with the
maximal observed energy $E_\text{max}$ we obtain,
\begin{align}
  a_S > 2.5\times 10^6 \,\text{cm} \times  \Big(\frac{d_S}{7\text{Gpc}}\Big)^2 
  \Big(\frac{f_{500}}{10^{-3}\text{sec}^{-1}\text{cm}^{-2}\text{keV}^{-1}}\Big)
  \Big(\frac{E_\text{max} }{1\text{MeV}}\Big)
  \Big(\frac{\varGamma}{1000}\Big)^{-5} \,,
  \label{agg}
\end{align}
where $f_{500}$ is the $\gamma$-ray flux at $500$\,keV and we have
assumed $\alpha=-2$, $z_S=1$ for the numerical estimate. 

Due to instrumental limitations the spectra of most GRBs are measured
up to $E_\text{max}\sim 1$~MeV. This corresponds to the energy $\hat
E\sim (1+z_S)\varGamma^{-1}\,\text{MeV}\ll 1\,\text{MeV}$ in the blob rest
frame. If the spectrum were cut at these energies, all the photons would be
below the threshold of pair production and we would not have
to worry about absorption at all. However, it is believed that the spectrum of
a typical GRB extends to much higher energies. This is supported by
detection of high energy emission (up to $\sim 100\,\text{GeV}$)
from several GRBs~\cite{Nava:2018qkq}. When such observations are
available, the corresponding maximal energy can be used in~\eqref{agg}
to set a strong lower bound on the source size. 

We now turn to absorption due to Compton scattering of $\gamma$-photons on
electrons and positrons. We will consider $e^{\pm}$ created by the process of
$\gamma\gamma$ scattering discussed above. As will be seen shortly, the
resulting constraint is stronger than (\ref{agg}) in the case when a direct
measurement of the high-energy component of $\gamma$-radiation is absent
\cite{2001ApJ...555..540L}. 

Assuming, for simplicity, that electrons are non-relativistic in the rest frame
of the emitting material, the optical depth for photons with $\hat E\lesssim
m_ec^2$ is essentially independent of their energy, 
\begin{align}
  \tau_{\gamma e}\simeq a_S\sigma_T\hat n_e\,,
\end{align} 
where $\hat n_e$ is the total density of electrons. The latter is
estimated as the density of photons for which the blob is optically
thick with respect to pair production,
\begin{align}
  \hat n_e = \int_{\hat\epsilon_\text{cr}}d\hat\epsilon\;\hat n_\gamma(\hat\epsilon)\,,
\end{align}
where $\hat \epsilon_\text{cr}$ is determined from the equation
$\tau_{\gamma\gamma}(\hat\epsilon_\text{cr})=1$. Using the expression
\eqref{eq:taugg2} we obtain,
\begin{align}
  \tau_{\gamma e}\simeq\frac{\eta(\alpha)}{-\alpha-1}\bigg[
  \frac{d_S^2\sigma_T}{c
    a_S}A\varGamma^{\alpha-2}(1+z_S)^{-\alpha-2}(m_ec^2)^{\alpha+1}\bigg]^2\;. 
  \label{tauge1}
\end{align}
Requiring that the emission region be optically thin, $\tau_{\gamma e}<1$,
translates into
\begin{align}
  a_S > \bigg(\frac{\eta(\alpha)}{-\alpha-1}\bigg)^{1/2}\frac{d_S^2\sigma_T}{c}
        \varGamma^{\alpha-2}(1+z_S)^{-\alpha-2}\,m_ec^2f(m_ec^2)\;.
  \label{age}
\end{align}
For the numerical values $\alpha=-2$, $z_S=1$ this yields,
\begin{align}
  a_S > 1.8\times 10^{9}\, \Big(\frac{d_S}{7\text{Gpc}}\Big)^2
        \Big(\frac{f_{500}}{10^{-3}\text{sec}^{-1}\text{cm}^{-2}\,\text{keV}^{-1}}\Big)
        \Big(\frac{\varGamma}{1000}\Big)^{-4}\text{cm} \,.
 \label{age1}
\end{align}
We see that for extreme values of the boost factor $\varGamma\sim 1200$ which
may occur in some GRBs~\cite{2011ApJ...738..138R} the lower bound on the source
size is quite small.  Of course, it would be too optimistic to interpret this
lower bound as a plausible value of $a_S$. On the other hand, the above
derivation is only an order-of-magnitude estimate. Moreover, it relies on a
power-law extrapolation of the GRB spectrum. As such, it shows that the
possibility for some GRBs to have sizes $a_S \lesssim 10^9$\;cm is not
completely excluded. Further progress in our quantitative understanding of the
physics of GRBs is required to conclude if this option is viable or
not.
Note that according to eq.~(\ref{eq:ahatestim}), if such GRBs exist, they
are expected to have variablity at sub-millisecond time scale.

\bibliography{lensing}
\bibliographystyle{JHEP}

\end{document}